\documentclass[prl,twocolumn,aps, secnumarabic,balancelastpage,amsmath,amssymb,superscriptaddress]{revtex4-1}

\usepackage{graphicx}
\usepackage{bm,amsmath, amssymb}
\usepackage{float}
\usepackage{comment}
\usepackage{tcolorbox}
\usepackage{todonotes}
\usepackage{makecell}
\usepackage{hyperref}
\usepackage{hhline}
\usepackage[utf8]{inputenc}

\newcommand{\ket}[1]{|#1\rangle}

\begin{document}
\title{Many-Body Chaos in the Sachdev-Ye-Kitaev Model}

\author{Bryce Kobrin}
\affiliation{Department of Physics, University of California, Berkeley, CA 94720, USA}
\affiliation{Materials Sciences Division, Lawrence Berkeley National Laboratory, Berkeley, CA 94720, USA}

\author{Zhenbin Yang} 
\affiliation{Department of Physics, Princeton University, Princeton, NJ 08540, USA}
\affiliation{Stanford Institute for Theoretical Physics, Stanford, CA, 94305, USA}

\author{Gregory D.~Kahanamoku-Meyer} 
\affiliation{Department of Physics, University of California, Berkeley, CA 94720, USA}

\author{Christopher T.~Olund} 
\affiliation{Department of Physics, University of California, Berkeley, CA 94720, USA}

\author{Joel E.~Moore} 
\affiliation{Department of Physics, University of California, Berkeley, CA 94720, USA}
\affiliation{Materials Sciences Division, Lawrence Berkeley National Laboratory, Berkeley, CA 94720, USA}

\author{Douglas Stanford} 
\affiliation{Stanford Institute for Theoretical Physics, Stanford, CA, 94305, USA}
\affiliation{Institute for Advanced Study, Princeton, NJ 08540, USA}

\author{Norman Y.~Yao}
\affiliation{Department of Physics, University of California, Berkeley, CA 94720, USA}
\affiliation{Materials Sciences Division, Lawrence Berkeley National Laboratory, Berkeley, CA 94720, USA}

\begin{abstract}
Many-body chaos has emerged as a powerful framework for understanding thermalization in strongly interacting quantum systems. 
While recent analytic advances have sharpened our intuition for many-body chaos in certain large $N$ theories, it has proven challenging to develop precise numerical tools capable of exploring this phenomenon in generic Hamiltonians. 
To this end, we utilize massively parallel, matrix-free Krylov subspace methods to calculate dynamical correlators in the Sachdev-Ye-Kitaev (SYK) model for up to $N=60$ Majorana fermions.
We begin by showing that numerical results for two-point correlation functions agree at high temperatures with dynamical mean field solutions, while at low temperatures finite-size corrections are quantitatively reproduced by the exactly solvable dynamics of near extremal black holes. 
Motivated by these results, we develop a novel finite-size rescaling procedure for analyzing the growth of out-of-time-order correlators (OTOCs).
Our procedure accurately determines the Lyapunov exponent, $\lambda$, across a wide range in temperatures, including in the regime where $\lambda$ approaches the universal bound, $\lambda = 2\pi /\beta$.
\end{abstract}

\maketitle
Characterizing thermalization in strongly interacting quantum systems is a goal that spans across multiple disciplines ranging from condensed matter and quantum information to quantum gravity.
Recent developments toward this goal have revealed striking insights into the relationship between quantum chaos and the delocalization, or scrambling, of quantum information. 
This unification is partly provided by the notion of out-of-time-order correlators (OTOCs), which take the general form $\left <W(t) V(0) W(t) V(0)\right>$ for local operators V and W \cite{Larkin_quasiclassical_1969,Shenker_black_2014,kitaevfundamental}. 
From an information theoretic perspective, these correlators determine the degree to which local information becomes hidden in nonlocal degrees of freedom, leading to the effective memory loss of initial conditions \cite{Hosur_chaos_2016,Shenker_black_2014}. 
From the perspective of chaos, OTOCs measure the sensitivity of one operator towards a small perturbation induced by another operator at an earlier time \cite{Maldacena_bound_2016,Shenker_stringy_2015}. 
In particular, for semiclassical chaotic systems, OTOCs are expected to exhibit a period of exponential growth analogous to the classical butterfly effect \cite{Liao_nonlinear_2018,LewisSwan_unifying_2019}.%

\begin{figure}[b!]
  \centering
  \includegraphics[width=\linewidth]{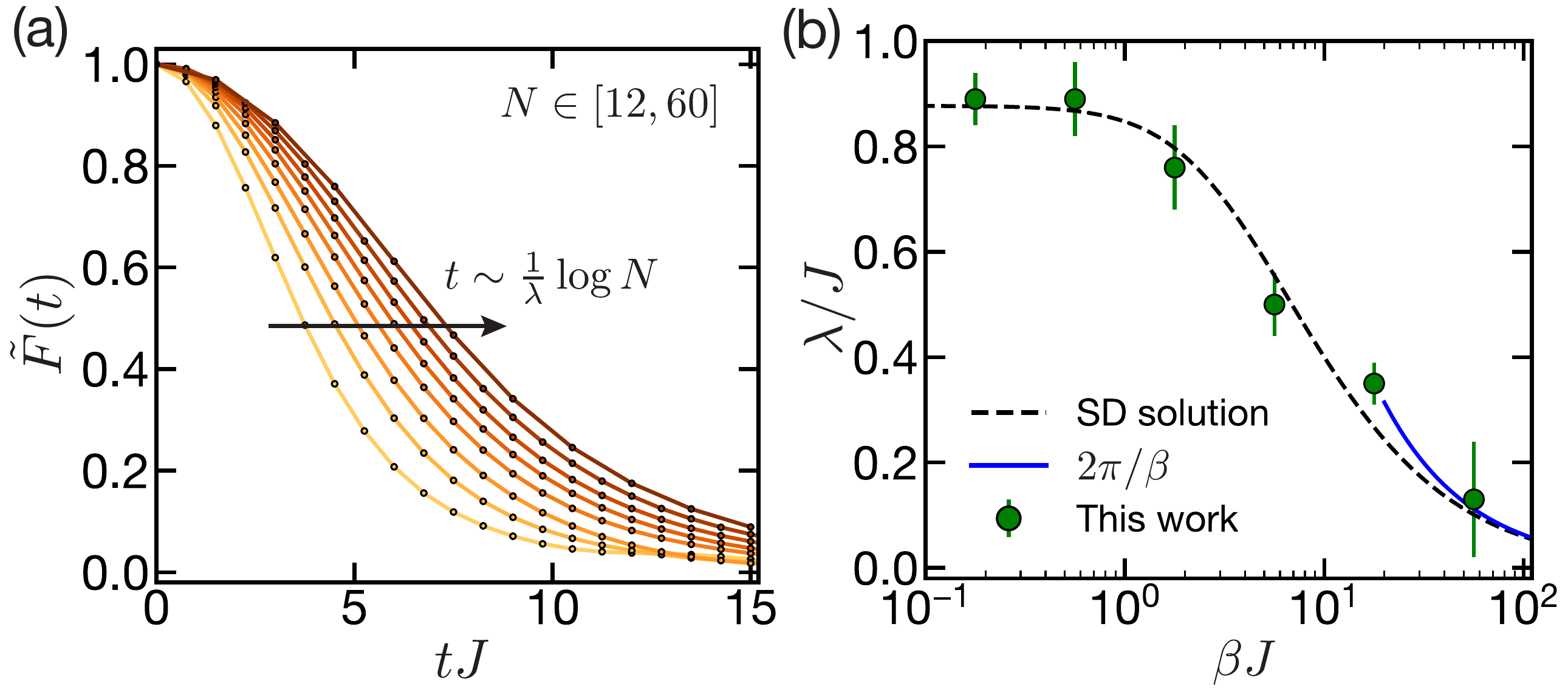}
  \caption{Regularized OTOCs in the SYK model, $\tilde F(t) \equiv F(t)/F(0)$, as shown for $\beta J = 10$ and system sizes $N \in [12,60]$. The early-time behavior is characterized by $1-\tilde F(t) \sim e^{\lambda t}/N$ and different system sizes are approximately related by a time translation symmetry, $t \rightarrow t + 1/\lambda \log N$. (b) Applying a finite-size rescaling procedure to the data, we determine $\lambda$ as a function of temperature (points). Our results exhibit excellent agreement with the theoretical predictions of the Schwinger-Dyson (SD) equations (dashed line), including in the regime where $\lambda$ approaches the bound on chaos $2\pi/\beta$ (blue).
    }
        \label{fig1}
\end{figure}

At the intersection between these two perspectives lies the discovery of a new form of quantum chaos in strongly interacting systems, known as many-body chaos.
This phenomenon is characterized by OTOCs whose leading order behavior is given by $e^{\lambda t}/N$, where $\lambda$ is the Lyapunov exponent and $N$ is related to the number of degrees of freedom per site \cite{Maldacena_bound_2016,Sekino_fast_2008}.
While such behavior was first anticipated in \cite{Sekino_fast_2008} and confirmed using holographic duality in \cite{Shenker_black_2014}, the first concrete Hamiltonian model to exhibit many-body chaos was introduced by Kitaev following previous work by Sachdev and Ye \cite{Sachdev_gapless_1993,Maldacena_remarks_2016,Kitaev_simple_2015,Kitaev_soft_2018}.
Remarkably, at low temperatures, the Lyapunov exponent of  this so-called SYK model saturates a  universal bound, $\lambda \le 2\pi T$, where $T$ is the temperature of the system \cite{Maldacena_bound_2016}.
The saturation of this bound is known to occur in theories of quantum gravity and their holographic duals \cite{Shenker_stringy_2015}, and indeed a direct correspondence has since been established between the low temperature dynamics of the SYK model and a universal theory of near extremal black holes (i.e.~Jackiw-Teitelboim gravity) \cite{Jensen_chaos_2016,Maldacena_conformal_1606,Engelsoy_an_2016,Kitaev_soft_2018}.
More recently, a number of other models that exhibit many-body chaos have been studied; however, their rate of chaos is parametrically slower than the thermodynamic bound \cite{Stanford_many-body_2016,Chowdhury_onset_2017}. 
In parallel, there have also been numerous proposals to directly measure OTOCs in coherently controlled quantum simulators  \cite{Joshi_quantum_2001,lantagne2020diagnosing, Yao_interferometric_1607,Zhu_measurement_2016,Swingle_measuring_2016,Li_measuring_2017,dressel2018strengthening,LewisSwan_unifying_2019,Bentsen_treelike_2019,Yoshida_disentangling_2019}, as well as a number of experimental demonstrations in small-scale systems \cite{Li_measuring_2017, Garttner_measuring_2017, Landsman_verified_2019,blok2020quantum}.

A major hurdle in benchmarking these experiments/proposals and in identifying novel models that exhibit many-body chaos, is the lack of a reliable numerical toolset.  
Indeed, in order to observe a period of clear exponential growth, the scrambling time must be well-separated from other effects related to local relaxation that occur at early times \cite{Maldacena_bound_2016,Bagrets_power-law_2017,Yao_interferometric_1607}

In this Letter, we take steps to overcome these challenges by employing massively parallelized Krylov subspace methods and developing new extrapolation tools to characterize many-body chaos. 
Specifically, we compute correlation functions for the SYK model for systems of up to $N=60$ Majorana fermions and leverage the model's correspondence with quantum gravity to interpret finite-size effects.
We present two main results.
First, we demonstrate that our numerical results for two-point functions, $G(t) = \left < W(t) W(0) \right >$, agree quantitatively with analytic predictions in two distinct regimes: (i)
at high temperatures, our results match the mean-field solution of the microscopic model, and 
(ii) at low temperatures, our results are consistent with the full quantum dynamics of near extremal black holes. 
These latter results represent, to the best of our knowledge, the first direct numerical verification of quantum gravity correlators, and highlight the close connection between finite-size corrections and gravitational fluctuations. 

Second, we introduce an extrapolation procedure for determining the Lyapunov exponent that explicitly takes into account higher-order terms in the OTOCs. 
We verify that this procedure accurately determines $\lambda$ as a function of temperature, including at low temperatures where $\lambda \approx 2\pi T$ (Fig.~\ref{fig1}).

\begin{figure}[t!]
  \centering
  \includegraphics[width=2.9in]{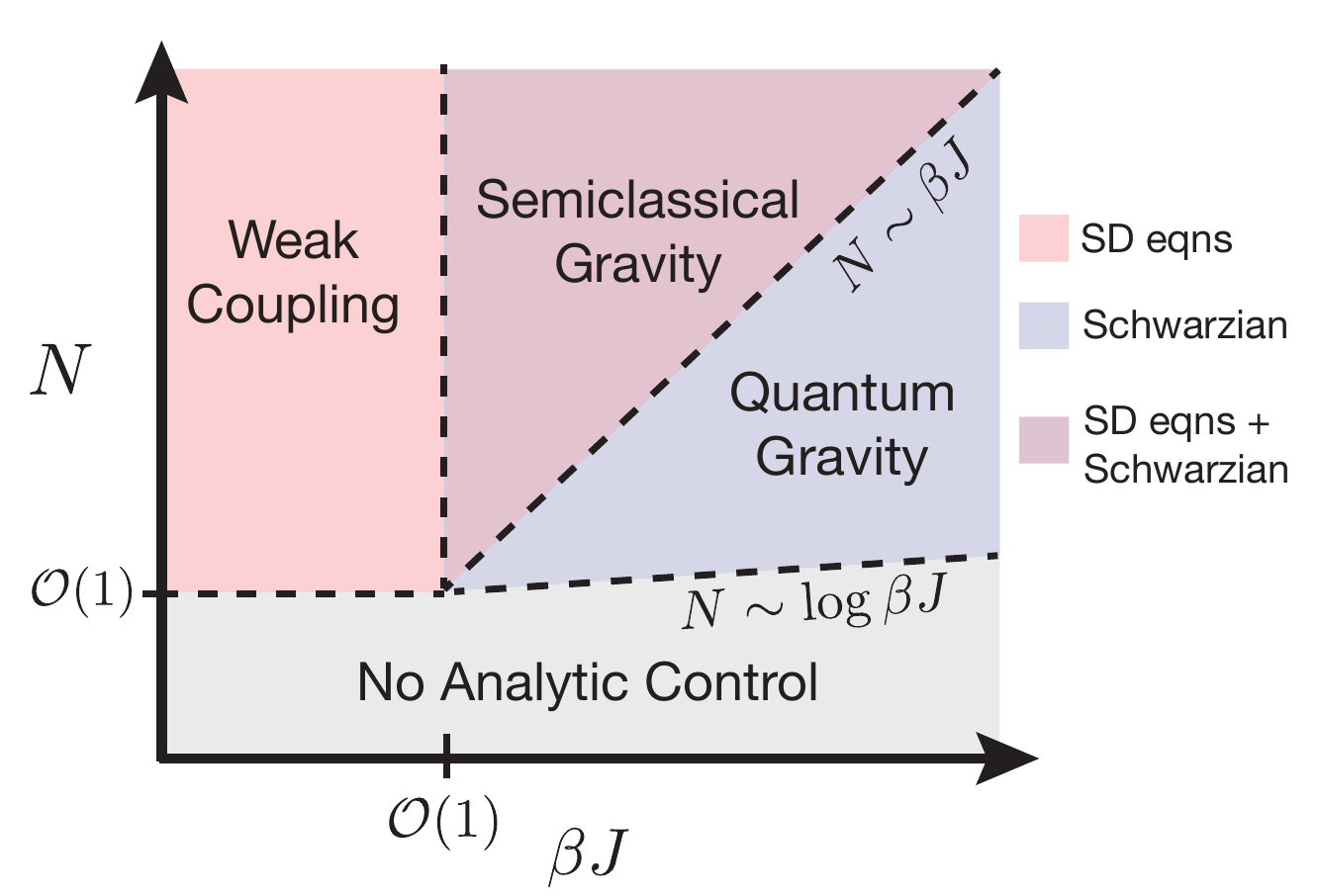}
  \caption{Regimes of analytic control for the SYK model as a function of system size, $N$, and inverse temperature, $\beta J$. In the semiclassical limit (red and purple), the model is well-described by a dynamical mean-field solution (Schwinger-Dyson equations). At low temperatures, finite-size corrections can be calculated using the Schwarzian action (blue), which is dual to AdS$_2$ gravity. However, at sufficiently small sizes (gray), the dynamics are governed by the discreteness of the energy spectrum and neither effective theory provides a valid description. 
    }
        \label{fig2}
\end{figure}

\emph{The SYK model and its gravity dual.---}
Consider the SYK Hamiltonian given by \cite{Kitaev_simple_2015,Maldacena_remarks_2016}:
\begin{equation}
H = \sum_{ i < j < k < l} J_{ijkl} \chi_i \chi_j \chi_k \chi_l .
\end{equation}
Here $\chi_i \; (i=1,\ldots,N)$ are Majorana fermions which obey the anti-commutation relation, $\left \{ \chi_i, \chi_j \right \} = \delta_{ij}$, and $J_{ijkl}$ are random (real) coefficients sampled from a Gaussian distribution with zero mean and variance $\overline{J_{ijkl}^2} = 6J^2/N^3$.

In order to probe the system's non-equilibrium dynamics, we will compute two different types of correlators. 
In-time correlators reveal how excitations in the system relax towards equilibrium.
In particular, we will consider the average imaginary-time Green's function, $G(\tau)$, and its real-time cousin, $G_R(t)$, given by
\begin{align} \label{G}
G(\tau) &\equiv \overline {\left < \chi_i(\tau) \chi_i(0) \right >}_\beta \\
G_R(t) &\equiv 2\textrm{Re}\left[\overline {\left < \chi_i(t) \chi_i(0)\right >}_\beta\right] 
\end{align}
where $\tau (t) >0$ is imaginary (real) time,  $\left < \cdots \right>_\beta = \frac {1}{Z}\textrm{Tr} \left [\cdots e^{-\beta H} \right ]$ is a thermal average at inverse temperature $\beta = 1/T$, and the overline denotes the (quenched) average over disorder realizations.
On the other hand, to probe chaos and  the scrambling of quantum information, we will consider out-of-time-order correlators.
We will primarily focus on the regularized OTOC,
\begin{equation} \label{F_reg}
F^{(\textrm{r})}(t) \equiv \overline {\left < \chi_i(t) \rho^{\frac 1 4} \chi_j(0) \rho^{\frac 1 4} \chi_i (t) \rho^{\frac 1 4} \chi_j (0) \rho^{\frac 1 4} \right > } 
\end{equation}
where $i \neq j$, and $\rho = e^{-\beta H}$, the imaginary-time evolution associated with the thermal ensemble, is distributed evenly among the four operators.
In the supplemental materials, we provide a detailed discussion regarding the key differences between this correlator and the unregularized version  \cite{regularize,supp}.

In the large $N$, semiclassical limit, both in-time and out-of-time correlators can be exactly computed via a diagrammatic approach \cite{Kitaev_simple_2015,Maldacena_remarks_2016}. 
The average Green's functions are determined by the self-consistent Schwinger-Dyson equations.
For the OTOCs, the leading order term in $1/N$ is computed by summing a series of  ladder diagrams.

Beyond the semiclassical limit, the dynamics at low temperature (i.e.~$\beta J \gg 1$) are captured by an effective theory known as the ``Schwarzian theory''  (Fig.~\ref{fig2}) \cite{Kitaev_soft_2018,Maldacena_remarks_2016,Bagrets_liouville_2016,Yang_quantum_2019}.
The same theory also describes Jackiw-Teitelboim gravity, a simple quantum gravity description of two-dimensional Anti-de-Sitter space.

Crucially, correlators in the Schwarzian theory are exactly computable \cite{Lam_shockwave_2018,Yang_quantum_2019}, which will enable us to perform quantitative, finite-size-scaling comparisons for two-point functions $G(\tau)$ and $G_R(t)$ \emph{outside} of the semiclassical limit. 
However, for the four-point function, the expressions are more complicated, and we will compare numerics to the ansatz: 
$
F(t) = C_0 + C_1\left(\frac {e^{\lambda t}} N \right)+C_2\left(\frac {e^{\lambda t}} N \right)^2 + \cdots
$, 
which is valid for large $N$ and $t\lesssim 1/\lambda \log N$ \cite{Maldacena_conformal_1606,Lam_shockwave_2018,Yang_quantum_2019}.
An analogous series expansion is expected to characterize OTOCs for the SYK model at high temperatures (and any other model described by ladder diagrams) \cite{lyapunov,Stanford_many-body_2016,Gu_relation_2019}.

\begin{figure}
  \centering
  \includegraphics[width=\linewidth]{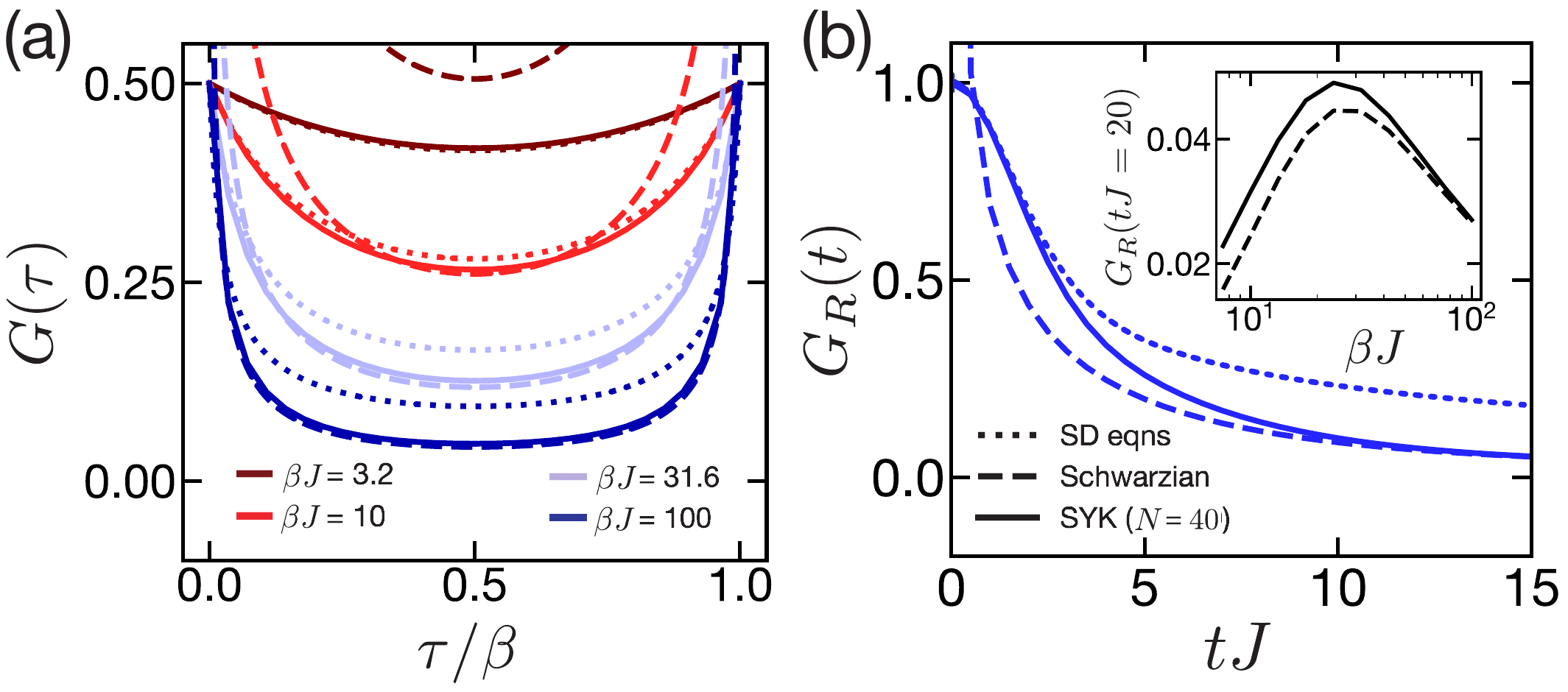}
  \caption{Two-point correlation functions in real and imaginary time. (a) Comparison of imaginary-time evolution between our numerics with 40 Majoranas (solid), the large $N$ solution (dotted), and the Schwarzian action (dashed). At high temperatures, we observe quantitative agreement between our numerical results and the large $N$ solution, while at low temperatures our numerics are well-described by the Schwarzian action. (b) Analogous comparison for real-time evolution with $\beta J = 56$. Our numerics show excellent agreement with the Schwarzian action for $t J \gtrsim 10$. The disagreement at earlier times is attributed to the difference in high-energy modes, which are cut off at the energy scale $J$ in the SYK model and are unbounded in the effective action. (inset) A salient feature in our real-time numerics is a non-monotonic trend with respect to temperature, as shown for $t J = 20$. This behavior is captured by the Schwarzian action (dashed) and can be understood as a consequence of the square root edge of the energy spectrum. 
    }
        \label{fig3}
\end{figure}

\emph{Non-equilibrium dynamics in the SYK model.---}
Our central numerical tool is a massively-parallelized implementation of a class of iterative methods known as Krylov subspace methods \cite{saad1992analysis,supp,dynamite}.
These methods approximate the time evolution of an initial state, $\ket {\psi(t)} = e^{-iHt} \ket {\psi}$, within a subspace formed by successive applications of the Hamiltonian.
Since this requires an initial pure state, we approximate thermal averages by taking the expectation value with respect to a Haar-random state $\ket {\tilde \psi}$ \cite{goldstein2006canonical,Steinigeweg_spin-current_2014,Luitz_ergodic_2017}:
\begin{equation}
\textrm{Tr} \left [ \hat O e^{-\beta H} \right ] \approx \left< \tilde \psi \right | e^{-\frac \beta 2 H} \hat O e^{-\frac \beta 2 H} \left| \tilde \psi \right >. 
\end{equation}
Owing to quantum typicality, the error in this approximation scales inversely with the  number of states in the thermal ensemble and, thus, decreases exponentially with $N$ for arbitrary systems and temperatures (above the spectral gap) \cite{Steinigeweg_spin-current_2014}.
In practice, we further reduce the error by averaging over initial states \cite{supp}.  

To begin probing the thermalizing dynamics of the SYK model, we compute the average Green's functions for both real- and imaginary-time evolution in the temperature range, $0 < \beta J \le 100$. 
At high temperatures, the imaginary-time Green's function, $G(\tau)$, shows  excellent agreement with the semiclassical solution given by the Schwinger-Dyson equations [Fig.~\ref{fig3}(a)]. 
At lower temperatures, the difference between our numerics and the semiclassical solutions widens. 
To understand the origin of these corrections, we plot the full solution predicted by the Schwarzian action. 
This exhibits close quantitative agreement with our data at temperatures corresponding to $\beta J \gtrsim 50$.
Crucially, this confirms that the Schwarzian action, or its corresponding gravity dual, accurately captures finite-size corrections away from the semiclassical regime. 

A few  remarks are in order. 
First, we note that the agreement with the Schwarzian action is only valid for system sizes larger than $N\approx 30$ \cite{supp}.
For smaller sizes, we observe additional finite-size corrections that are attributed to the discreteness of the energy spectrum. 
Such non-Schwarzian corrections are expected to dominate when the temperature approaches the energy of the level spacings (i.e.~$N \sim \log \beta J$ in Fig.~\ref{fig2}) \cite{Georges_quantum_2001,GurAri_does_2018}.
Second, the agreement between the Schwarzian and our numerics does not hold at timescales shorter than the inverse of the microscopic coupling strength (i.e.~$\tau J \lesssim 1$); specifically, the Schwarzian dynamics diverge as $\tau J \rightarrow 0$ while our numerics approach a finite value.
This difference arises from the fact that the Scharzian action is the effective theory only at low energies (compared to $J$); for higher energies, the SYK dynamics are governed by the microscopic nature of the model.

Much like the imaginary-time case, we find that the retarded Green's function, $G_R(t)$,  agrees with the semiclassical solutions at high temperatures and with the full dynamics of the Schwarzian action at low temperature [Fig.~\ref{fig3}(b)]. 
We note, however, that the early-time discrepancy with the Schwarzian action is extended to later times (i.e.~$\beta J \sim 10$).
This can be attributed to the longer timescale required for the phase cancellation of the high-energy modes in real time, as opposed to the direct suppression that occurs in imaginary time. 

Working with real-time dynamics also allows us to probe a rather non-trivial prediction of the Schwarzian action. 
In particular, one expects  the late-time dynamics to be governed by the functional form of the spectral density at low energies, $\rho(E) \sim E^{\frac 1 2}$ \cite{GarciaGarcia_analytical_2017,Bagrets_liouville_2016,Cotler_black_2017}.
This square-root singularity leads to a power-law decay of the Green's function, with a power that depends on both the temperature and the timescale.
Intriguingly, it predicts a \emph{non-monotonic} temperature dependence for the decay of the Green's function, in stark contrast to the monotonic dependence predicted by the semiclassical solution. 
This non-trivial temperature dependence, consistent with only the full Schwarzian solution, is indeed borne out by the numerics [inset, Fig.~\ref{fig3}(b)]. 

\emph{Lyapunov exponent of  the SYK model.---}To probe many-body chaos in the SYK model, we now compute regularized OTOCs [Eqn.~\ref{F_reg}] for temperatures in the range $0 < \beta J \le 56$ and for system sizes up to $N=60$.
In the large $N$ limit, one expects a well-defined period of exponential growth, starting from the timescale at which the two-point functions decay and persisting until the scrambling time \cite{Maldacena_bound_2016}.
However, for conventional exact diagonalization studies, there is little separation between these timescales, owing to the limited system sizes that are numerically accessible; indeed, prior studies actually observed an increase in the extracted Lyapunov exponent as a function of decreasing temperature --- the opposite behavior of what is expected \cite{,Fu_numerical_2016,supp}. 
By scaling to larger system sizes using Krylov subspace methods, we observe a direct turnover in this trend. Moreover, we introduce a novel extrapolation method, which provides a robust way of extracting the Lyapunov exponent. 

The intuition behind our method is as follows:
For a large class of many-body chaotic systems, the full form of the OTOC in the semiclassical limit is given by a series in $e^{\lambda t}/ N$.
Crucially, this series exhibits a rescaling symmetry, wherein $N \rightarrow r N$ amounts to shifting the full curve by $t \rightarrow t + 1/\lambda \log r$.
This symmetry can be shown explicitly for the Schwarzian action, which governs low-temperature dynamics of the SYK model, and is also expected to hold at high temperatures \cite{Stanford_many-body_2016,Gu_relation_2019}. 

This suggests that we can determine $\lambda$ at a given temperature by attempting to collapse our data [Fig.~\ref{fig1}(a)] through finite-size rescaling of the form $t \rightarrow t + 1/\lambda \log N$.
More specifically,  we first interpolate our data to find the time, $t_*$, at which each curve crosses a fixed value, i.e.~$F(t_*)/F(0) = 1-F_0$ \cite{supp}.
Next, we estimate $\lambda_{\textrm{fit}}(N)$ as $1/\lambda_{\textrm{fit}} = dt_*/d(\log N)$, where $N$ corresponds to the system size about which we take the numerical derivative.
Finally, we fit our results to a $1/N$ series, $\lambda_{\textrm{fit}}(N) = \lambda_0 + \lambda_1 / N + \lambda_2 / N^2 + \cdots$; the leading order term $\lambda_0$ corresponds to the extrapolated value for $\lambda$ as $N \rightarrow \infty$.

In Fig.~\ref{fig1}(b), we present our results for $\lambda_0$ as a function of temperature. 
We observe excellent agreement with analytic predictions for all temperatures in the range $0 < \beta J \le 56$.
Crucially, our protocol works at \emph{low-temperatures} where the $2\pi/\beta$ scaling (saturating the bound on chaos) becomes apparent. 

An important question to ask is over what range of temperatures we expect our procedure to remain valid. 
There are three relevant considerations.
First,  the temperature must be high compared to the energy associated with the level spacing; we account for this requirement by considering only system sizes where at least 20 eigenstates, on average, lie within $\Delta E = 1/\beta$ of the ground state.
Second, the system must be sufficiently close to the semiclassical limit for the rescaling symmetry to hold. 
It is known from the Schwarzian action that this condition corresponds to $\beta J \lesssim N$.
Asymptotically this is a much stronger requirement than the first condition; however, for the system sizes relevant for our study ($N \lesssim 60$) both requirements imply a low temperature limit of $\beta J \approx 60$. 

Third, there must sufficient separation between the scrambling time and the short-time dissipative dynamics. 
In the case of the regularized correlator, this condition is given by $\beta J \lesssim N$, leading to the same temperature range as the semiclassical requirement. 
However, in the case of unregularized correlators, the corresponding condition is $(\beta J)^3 \lesssim N$; this implies that the unregularized correlator is subject to stronger finite-size effects, which we corroborate through our numerics \cite{supp}.

\emph{Discussion and outlook.---}By employing massively parallelized Krylov subspace methods and developing novel extrapolation tools, we have demonstrated that one can utilize numerics to accurately capture the thermalizing and chaotic dynamics of the SYK model.
Our results for two-point Green's functions represent a direct verification of the dynamics of quantum black holes in a highly fluctuating regime.
Moreover, our finite-size rescaling procedure for extracting Lyapunov exponents leads to the first numerical evidence that the SYK model saturates the theoretical bound on chaos, $\lambda \approx 2\pi T$.

We anticipate that the numerical tools demonstrated here will open the door to a number of intriguing future directions. 
First, our numerical tools can be applied to variations of the SYK model (i.e.~large $q$ limit) for which the effective action (i.e.~Liouville action) is known for all temperatures \cite{Cotler_black_2017,Erdos_phase_2014,Berkooz_towards_2019}. 
This will enable quantitative studies of finite-size corrections in the high-temperature regime, where the Schwarzian action is not valid. 
Second, our procedure for characterizing Lyapunov exponents can diagnose many-body chaos in other models beyond the SYK model;
this is of particular relevance for experimental platforms which have constraints on the types of interactions and disorder that can be realized \cite{Chew_approximating_2017,Danshita_creating_2017,Bentsen_treelike_2019,Landsman_verified_2019,blok2020quantum}.
Finally, we envision future numerical simulations to test more complex gravitational phenomena, including traversable wormholes \cite{Gao_traversable_2017,Maldacena_diving_2017}, and the possible emergence of SYK dynamics in transport experiments of quantum materials \cite{kruchkov2019thermoelectric, chen2018quantum, altland2019sachdev}.

We gratefully acknowledge the insights of and discussions with Felix Flicker, Snir Gazit, Thomas Scaffidi, Pratik Rath, Markus Schmitt, Nicole Yunger Halpern, Brian Swingle, Victor Galitski, Stephan Plugge, and Beni Yoshida.
This work was supported by the U.S. Department of Energy through the Quantum Information Science Enabled Discovery (QuantISED) for High Energy Physics (KA2401032) and through the GeoFlow Grant No. de-sc0019380.
This research used resources of the National Energy Research Scientific Computing Center (NERSC), a U.S. Department of Energy Office of Science User Facility operated under Contract No. DE-AC02-05CH11231. 
  The numerical work performed in this work used the \texttt{dynamite} Python frontend \cite{dynamite}, which supports a matrix-free implementation of Krylov subspace methods based on the \texttt{PETSc} and \texttt{SLEPc} packages \cite{Hernandez__0509, Roman__16, Balay__17}.
J.E.M. acknowledges support of NSF DMR-1918065 and a Simons Investigatorship.
G.D.K.-M. acknowledges support by the Department of Defense (DoD) through the National Defense Science \& Engineering Graduate Fellowship (NDSEG) Program.
C.O. acknowledges support by the Department of Defense through the National Defense Science \& Engineering Graduate Fellowship program.

\bibliography{ref_auto.bib,ref_manual.bib}

\end{document}


\title{Supplementary Information:\\ Many-Body Chaos in the Sachdev-Ye-Kitaev Model}

\author{Bryce Kobrin}
\affiliation{Department of Physics, University of California, Berkeley, CA 94720, USA}
\affiliation{Materials Sciences Division, Lawrence Berkeley National Laboratory, Berkeley, CA 94720, USA}

\author{Zhenbin Yang} 
\affiliation{Department of Physics, Princeton University, Princeton, NJ 08540, USA}
\affiliation{Stanford Institute for Theoretical Physics, Stanford, CA, 94305, USA}

\author{Gregory D.~Kahanamoku-Meyer} 
\affiliation{Department of Physics, University of California, Berkeley, CA 94720, USA}

\author{Christopher T.~Olund} 
\affiliation{Department of Physics, University of California, Berkeley, CA 94720, USA}

\author{Joel E.~Moore} 
\affiliation{Department of Physics, University of California, Berkeley, CA 94720, USA}
\affiliation{Materials Sciences Division, Lawrence Berkeley National Laboratory, Berkeley, CA 94720, USA}

\author{Douglas Stanford} 
\affiliation{Stanford Institute for Theoretical Physics, Stanford, CA, 94305, USA}
\affiliation{Institute for Advanced Study, Princeton, NJ 08540, USA}

\author{Norman Y.~Yao}
\affiliation{Department of Physics, University of California, Berkeley, CA 94720, USA}
\affiliation{Materials Sciences Division, Lawrence Berkeley National Laboratory, Berkeley, CA 94720, USA}

\maketitle
\tableofcontents

\section{Numerical methods}

\subsection{Jordan-Wigner transformation}
To represent the SYK Hamiltonian numerically, we map $N$ Majorana operators into $N/2$ Pauli spin-1/2 operators using the canonical Jordan-Wigner transformation:
\begin{equation}
\chi_i \rightarrow \left\{\begin{array}{lr}
        \left (\prod_{j < k} \sigma_j^z\right) \sigma^x_{k}, \quad i \text{ even}\\
        \left (\prod_{j < k} \sigma_j^z\right) \sigma^y_{k}, \quad i \text{ odd}
        \end{array}\right .
\end{equation}
where $0 \le i < N$ is the Majorana index and $k = \text{floor}(i/2)$.
We note that the SYK Hamiltonian (Eq.~1 of the main text) contains a single $Z_2$ symmetry, corresponding to the conservation of fermionic charge parity, i.e.~$P = \left(\sum_i \sigma_i^z\right)\mod 2$ \cite{Cotler_black_2017}.
To take advantage of this symmetry, we prepare initial states in one of the symmetry sectors and evolve under the relevant sector of the Hamiltonian.
We verify that this simplification has a negligible effect on correlation functions compared to initial states that span both symmetry sectors. 

\begin{table} 
\begin{tabular}{c|c|c|c|c}
  \hline \hline
  \makecell{System \\ size ($N$)}  &  Hardware & \makecell{Memory,\\ matrix-free} & \makecell{Memory,\\ full-matrix} & \makecell{Time \\ per curve \\ (10 points)} \\ \hline
  \makecell{$20$} &    &  100 KB & 60 MB  & $<$1 sec \\
  \makecell{$30$} &  GPU  &  2 MB & 10 GB   & 20 sec \\
  \makecell{$40$} &  (Nvidia V100) & 40 MB & 1 TB & 30 min \\ 
  \makecell{$50$} &  & 1 GB & 100 TB & 20 hrs \\ \hline
  \makecell{$60$} & \makecell{CPU (Intel KNL) \\ 512 nodes} & 40 GB & 6 PB & 70 hrs\\
  \hline \hline
\end{tabular}
\caption{Summary of computational requirements for computing OTOCs using Krylov subspace methods \cite{dynamite}. All computations in this work were performed using matrix-free methods; for comparison, we also list the memory requirement for standard, full-matrix computations.  } \label{table:computation}
\end{table} 

\subsection{Krylov subspace methods}
The computational workhorse in our study is a class of iterative methods known as Krylov subspace methods \cite{saad1992analysis,Roman__16}.
These methods approximate the action of the unitary operator $U(t) = e^{-iHt}$ by projecting $H$ onto a smaller subspace.
This subspace  -- the so-called Krylov subspace -- is formed by successively multiplying an initial state $\ket{\psi}$ by the Hamiltonian: $\left \{ \ket {\psi}, H \ket {\psi}, H^2 \ket {\psi}, \cdots, H^m \ket {\psi} \right \}$, where $m \sim \mathcal{O}(1)$ is the dimension of the subspace.
Time evolution of the initial state is then approximated by $e^{-i H t}\ket{\psi} \approx p_m(H) \ket{\psi}$, where $p_m(H)$ is a polynomial of degree $m$ and is determined by exponentiating the projection of $H$ within the subspace (whose dimensions are $m\times m$). 

The key advantage of this approach is that its core computational component is matrix-vector multiplication, a significantly more efficient task than exact diagonalization (ED).
For example, for the SYK Hamiltonian with $N$ Majoranas, the time complexity of matrix-vector multiplication (using sparse matrix techniques) scales as $\mathcal{O}(N^4 2^{N/2})$, whereas that of exact diagonalization is $\mathcal{O}(2^{3N/2})$.
Moreover, matrix-vector operations can take advantage of two important high-performance computing techniques.
First, they can be distributed across many parallel processes, thereby reducing the workload of each invididual processor.
Second, the memory cost can be substantially reduced using matrix-free methods: Instead of storing the Hamiltonian in matrix form, one computes the action of $H \ket \psi$ ``on the fly'' using a symbolic representation of $H$, which contains only $\mathcal{O}(N^4)$ terms.
In particular, we estimate the memory requirement with and without matrix-free methods as:
\begin{equation}
\begin{split}
\textrm{matrix-free} \; &=(\textrm{dimension of Hilbert space})\times(\textrm{number of Krylov vectors})\times (\textrm{bytes per element}) \\ & \quad \quad \quad \quad \quad \quad \quad \quad+ (\textrm{number of terms in }H)\times(\textrm{bytes per coefficient}) \\ &= 2^{N/2-1} \times 5 \times (16 \textrm{ bytes}) + {N \choose 4} \times (20 \textrm{ bytes}) \\ 
\end{split}
\end{equation}
\begin{equation}
\begin{split}
\textrm{full-matrix} \; &=(\textrm{dimension of Hilbert space})\times(\textrm{number of Krylov vectors})\times (\textrm{bytes per element}) \\ &  \quad + (\textrm{dimension of Hilbert space}) \times (\textrm{number of terms in }H)\times(\textrm{bytes per coefficient}) \\ &= 2^{N/2-1} \left [ 5 \times (16 \textrm{ bytes}) + {N \choose 4} \times (24 \textrm{ bytes}) \right ]
\end{split}
\end{equation}

In Table \ref{table:computation}, we summarize the computational requirements of our simulations and provide details on the specific hardware used in our study.
Crucially, the substantial memory savings offered by matrix-free methods allowed us to run the majority of our computations (up to $N = 50$) on GPUs, with a run time of at most one day per disorder average.
By contrast, prior to implementing matrix-free methods, the largest systems we were able to simulate were $N \approx 40$ due to the intensive memory requirements.
All of our numerics were performed with a custom open-source Python package called \texttt{dynamite} \cite{dynamite}, which provides a frontend interface for PETSc and SLEPc, two standard libraries for paralellized linear algebra computations \cite{Hernandez__0509, Roman__16, Balay__17}.

\begin{figure}[t!]
  \includegraphics[width=0.4\linewidth]{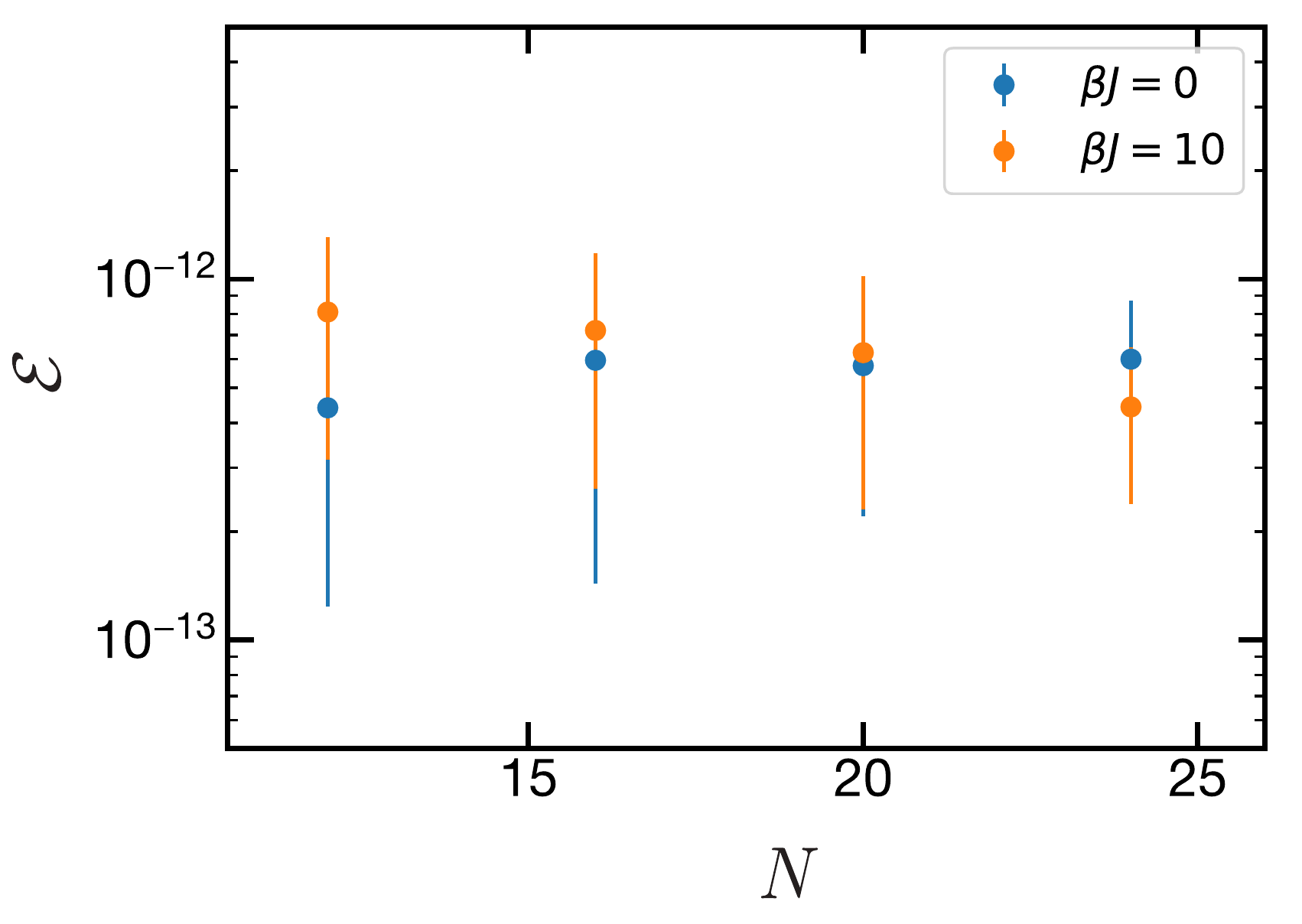}
  \caption{The numerical error of OTOCs computed with Krylov subspace methods compared to exact diagonalization. The error is quantified as the maximum difference in the OTOCs for timescales $tJ \le 20$ (Eq.~\ref{eq:error}). Error bars represent the standard deviation over 100 disorder realizations.}
        \label{fig:benchmarking}
\end{figure}

We note that there are two potential drawbacks with Krylov subspace techniques.
First, as an approximate technique, they introduce small numerical errors compared to exact evolution; nevertheless, these errors are well-controlled by working with a sufficiently large subspace and dividing the time evolution into a series of small time steps.
To estimate the magnitude of errors present in our study, we compute the absolute difference of out-of-time-order correlators (OTOCs) using Krylov subspace methods and exact diagonalization,
\begin{equation}\label{eq:error}
\mathcal{E} = \max_t \left| F_\textrm{ED}(t) - F_\textrm{Krylov}(t) \right| .
\end{equation}
%
The results are shown in Fig.~\ref{fig:benchmarking} for system sizes $N \in [12,24]$ and times $t J \in [0,20]$.
In all cases, the absolute error is less than $10^{-12}$ and is thus neglected for the rest of this study.
Second, and more substantially, the computational time scales (approximately linearly) with the evolution time.
Thus, Krylov subspace methods are best suited for evolving to intermediate timescales (e.g.~$tJ \lesssim 100$ where $J$ is the typical coupling strength), which is the case for all correlators considered in this study.


\subsection{Disorder averaging}
As discussed in the main text, we approximate thermal quantities by taking expectation values with respect to Haar-random initial states \cite{goldstein2006canonical,Steinigeweg_spin-current_2014,Luitz_ergodic_2017}: 
\begin{equation}\label{eq:q_typicality}
\langle \hat O \rangle_\beta = \frac{\textrm{Tr} \left [ \hat O e^{-\beta H} \right ]}{\textrm{Tr} \left [ e^{-\beta H} \right ]} \approx \frac{\left< \tilde \psi \right | e^{-\frac \beta 2 H} \hat O e^{-\frac \beta 2 H} \left| \tilde \psi \right >}{\left< \tilde \psi \right | e^{- \beta H}  \left| \tilde \psi \right >} .
\end{equation}
The error in this approximation is expected to decrease exponentially with $N$; in particular, at infinite temperature the error scales inversely with the dimension of the Hilbert space, while at finite temperature it scales with the number of states in the thermal ensemble, i.e.~$\textrm{Tr}[e^{-\beta(H-E_0)}]$ where $E_0$ is the ground-state energy \cite{Steinigeweg_spin-current_2014}.
This scaling follows from the concept of quantum typicality and is applicable to generic interacting systems (e.g.~as previously demonstrated for a Heisenberg spin chain \cite{Steinigeweg_spin-current_2014}). 
 
We generate Haar-random states by drawing each (complex) element in $\ket {\tilde \psi}$ from a random Gaussian distribution.
In practice, we average simultaneously over the initial states and disorder realizations of the coefficients in the Hamiltonian, $J_{ijkl}$.
Both forms of sampling -- the Hamiltonian coefficients $J_{ijkl}$ and the random initial state $\ket {\tilde \psi}$ -- give rise to numerical fluctuations which decrease significantly with system size.
For the former, the SYK model is self-averaging, implying that as $N \rightarrow \infty$ the correlation functions for a \emph{single} disorder realization approach the disorder average; specifically, one expects these fluctuations to decrease as a function of the number of random $J_{ijkl}$ coefficients, i.e.~polynomially with system size. 
For the latter, our method for approximating thermal averages with random pure states (Eq.~\ref{eq:q_typicality}) is expected to be accurate up to exponential corrections in the system size.

\begin{figure}[t!]
  \includegraphics[width=.9\linewidth]{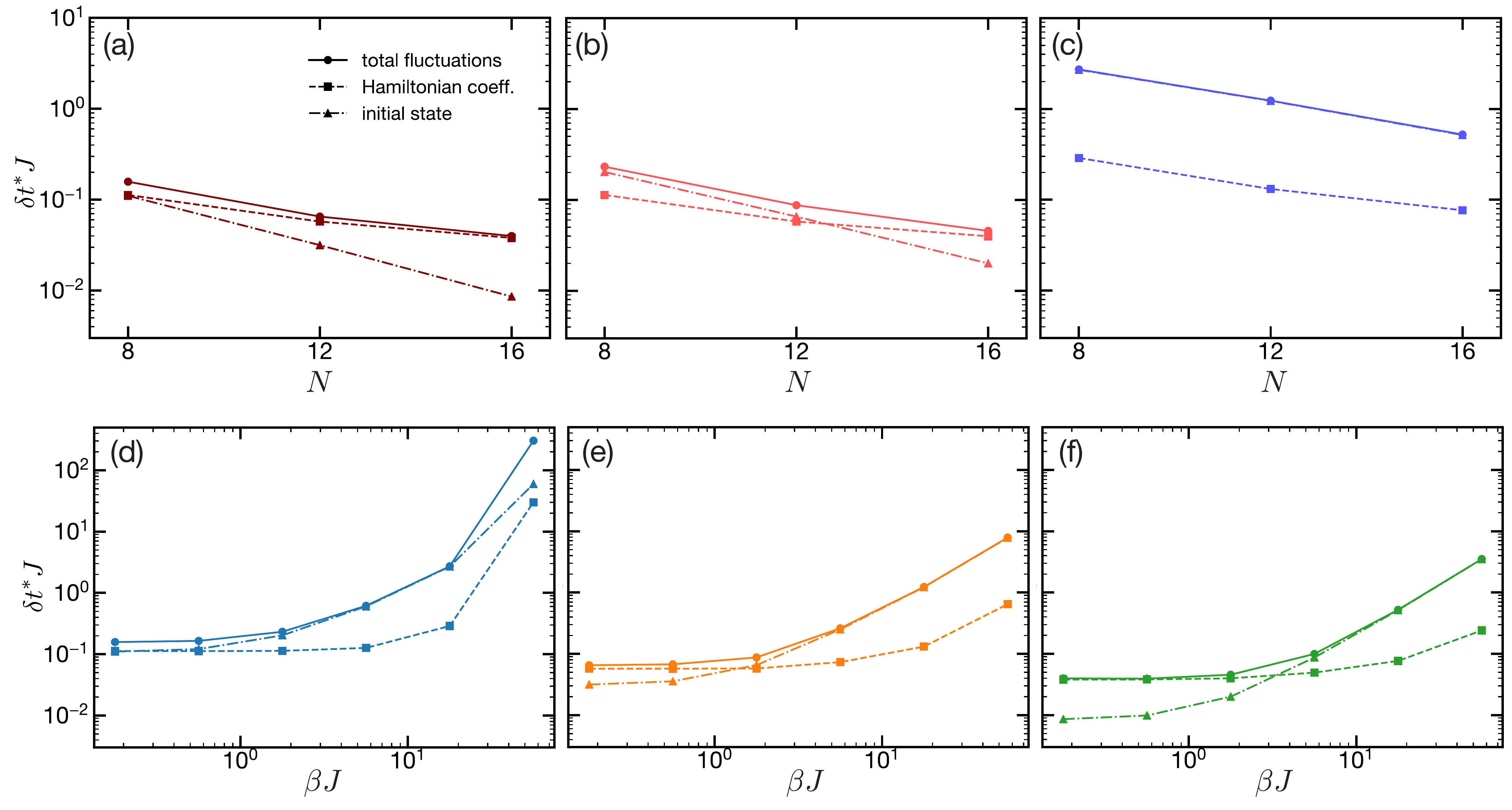}
  \caption{Disorder fluctuations of the OTOCs, as measured by the standard deviation of $t^*$ given by $1 - \tilde F(t^*) = 0.25$. Two sources of disorder contribute to the total fluctuations (solid): the Hamiltonian coefficients $J_{ijkl}$ (dashed) and the initial random state (dot-dashed). (a-c) Fluctuations vs. system size for (a) $\beta J=0.18$, (b) $\beta J=1.8$, and (c) $\beta J=18$. (d-f) Fluctuations vs. temperature for (d) $N=16$, (e) $N=24$, and (f) $N=32$.
 }
        \label{fig:fluctuations}
\end{figure}

To test these expectations, we compute the magnitude of numerical fluctuations with respect to each type of disorder.
In particular, we calculate the fluctuations in the timescale $t^*$ given by $\tilde F(t^*) = 0.25$ in two different ways:
(a) by fixing $J_{ijkl}$ and calculating the standard deviation with respect to different initial states; and (b) by averaging first over initial states and then determining the standard deviation with respect to different realizations of $J_{ijkl}$.
These results are shown in Fig.~\ref{fig:fluctuations}.
In general, both types of fluctuations are on the same order of magnitude, and their magnitude increases dramatically at small system sizes and low temperatures.
The size dependence is consistent with the aforementioned self-averaging behavior, while the temperature dependence is attributed to the reduced number of states that contribute to the behavior of the low-temperature correlators.

Although both sources of error decrease with system size, we emphasize that at the system sizes relevant for our study, both types of fluctuations are significant and extensive disorder averaging is required to obtain precise results (e.g.~we perform hundreds of disorder realizations even for $N \approx 40$ Majoranas). 
As a result, while single curves have been obtained for 60 Majoranas, the primary results for this study were based on $N \le 46$ Majoranas, for which sufficient disorder averaging could be performed. 

\section{Characterizing the Lyapunov exponent}

\subsection{Fitting to a simple exponential}
Several prior studies of many-body chaos have characterized Lyapunov exponents by fitting OTOCs to a simple exponential form, $\sim e^{\lambda t}$ \cite{Yao_interferometric_1607,lantagne2020diagnosing,Shen_out-of-time-order_2017,Keles_scrambling_2019}.
In this section, we apply this fitting procedure to our numerical data and compare our results to the known theoretical values for $\lambda$. 

\begin{figure}[b!]
  \includegraphics[width=0.65\linewidth]{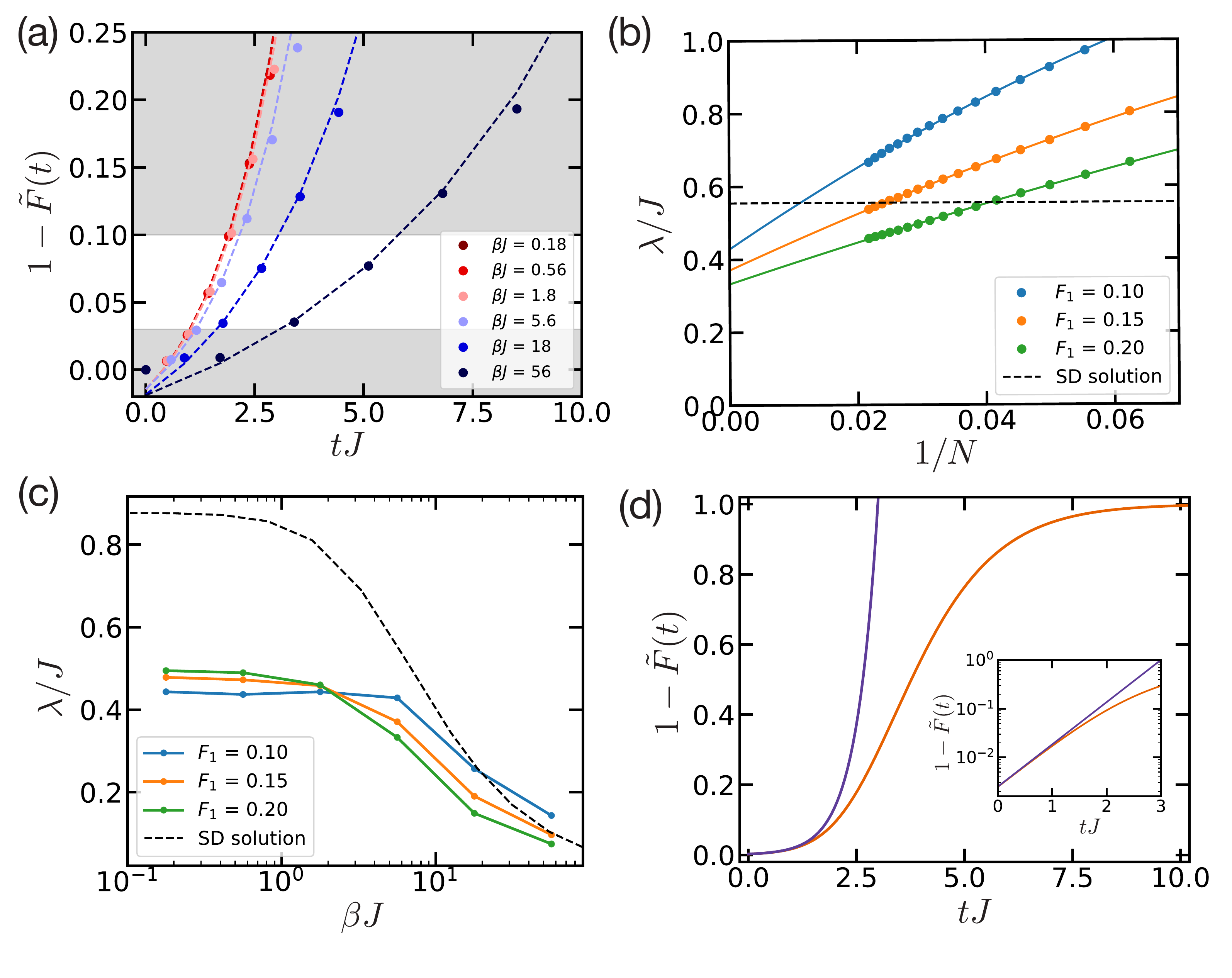}
  \caption{Extracting $\lambda$ by fitting to a simple exponential. (a) For each set of data, we apply least squares regression based on the fitting function, $a+b e^{\lambda_\textrm{fit} t}$, (dashed line) within a window $F_0 \le F(t) \le F_1$ (white area). (b) We then extrapolate $\lambda_\text{fit}$ as a function of system size through quadratic function in $1/N$ (solid lines). The results are shown for $\beta J = 5.6$, $F_0 = 0.03$, and various values for $F_1$. The dashed line represents the theoretical prediction for $\lambda$. (c) Extrapolated results for $\lambda$ as a function of temperature for several values of $F_1$; these exhibit significant disagreement with the theoretical results (dashed line). (d) The theoretical curve for $\tilde F(t)$ at low temperatures, given by \eqref{hypergeo}, (orange) compared to the leading-order simple exponential (purple). 
 }
        \label{fig:exp_fit}
\end{figure}

More specifically, we perform least squares regression on each regularized OTOC curve, $\tilde F(t) \equiv F(t)/F(0)$, using the fitting function, $a+b e^{\lambda_\textrm{fit} t}$, within a range defined by $F_0 \le 1-\tilde F(t) \le F_1$.
We then extrapolate the fits at different system sizes using a quadratic extrapolation, $\lambda_\textrm{fit}(N) = \lambda_0 + \lambda_1 / N + \lambda_2 /N^2$. 
In Fig.~\ref{fig:exp_fit}, we illustrate this procedure and show the extrapolated values for $\lambda$ using various values for $F_0$ and $F_1$. 
It is clear that this approach does not converge to the exact values of $\lambda$ for any temperature; indeed, the estimated values are approximately a factor of 2 smaller than the theoretical expectations. 

A few remarks are in order.
First, our fitting procedure differs slightly from other studies in the sense that we perform the fits for a fixed range in the (normalized) magnitude of the OTOCs rather than a fixed range in time, i.e.~$t_0 \le t \le t_1$.
We chose this approach because the growth of OTOCs occurs at different timescales, depending on the temperature and system size. 
To compare directly to previous work, we also tried fitting our data across a fixed range in time and found no improvements in the estimates for $\lambda$.
Second, we observe that our fitting results depend sensitively on the choice of $F_0$ and $F_1$, though for all choices of these parameters our results for $\lambda$ were inconsistent with theoretical expectations. 

In principle, one expects the best results using $F_1 \ll 1$, as the simple exponential form is only defined for the initial growth.
For a more precise estimate of the range of validity, we turn to the semiclassical solution for $F(t)$ at low temperatures, given by \eqref{hypergeo}, which takes into account higher order terms. 
By plotting this full solution against the leading order exponential, we find that the exponential is a good approximation only for $F_1 \lesssim 0.05$. 
In finite-size numerics, probing such small magnitudes presents several challenges.
First, the absolute size of numerical fluctuations is approximately constant at all times, and thus the relative size of fluctuations compared to the signal is enhanced for small $F_1$.
Second, one requires the timescale for which the exponential reaches $F_1$ to be much longer than the dissipation time. 
As the former timescale scales as $\log F_1 N$ and the latter timescale is constant, achieving this separation becomes more difficult as $F_1$ decreases. 
These challenges, as well as the poor results of our exponential fits, underscore our motivation for developing a fitting procedure that takes into account higher-order terms.

\subsection{Fitting to the low-temperature, semiclassical solution}

We next consider fitting our numerical data to semiclassical form of the OTOC at low temperatures (see Section below) \cite{Maldacena_conformal_1606}:
\begin{equation} \label{hypergeo}
\frac {F(t)}{F(0)} = \frac {U(2\Delta,1,\frac 1 z)}{z^{2\Delta}} , \quad z = \frac {e^{\lambda_\textrm{fit} t}}{N_\textrm{fit}}
\end{equation}
where $U$ is the confluent hypergeometric function, $\Delta = 1/4$ (i.e.~the conformal dimension), and we include two fitting parameters -- $\lambda_\text{fit}$ and $N_\textrm{fit}$ -- associated with the Lyapunov exponent and the system size, respectively.  
This function provides a phenomenological model for capturing higher-order effects that occur after the initial exponential growth (i.e.~saturation behavior). 
Nevertheless, we emphasize that the exact form of the function is only rigorously justified at low temperatures where the SYK model is described by the Schwarzian action. 

\begin{figure}[t!]
  \includegraphics[width=\linewidth]{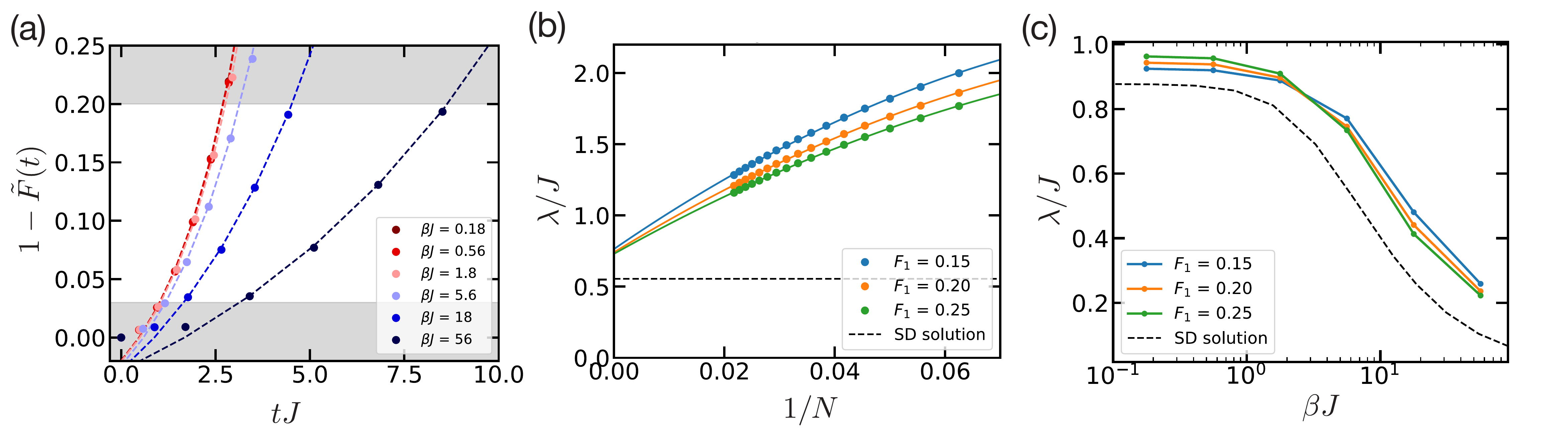}
  \caption{Extracting $\lambda$ by fitting to the low-temperature, semiclassical solution, i.e.~\eqref{hypergeo} with $\Delta = 1/4$. Analogous to Fig.~\ref{fig:exp_fit}, we fit each set of data using least squares regression (a) and perform a $1/N$ extrapolation on the best-fit values for $\lambda$ (b), as shown for $\beta J = 5.6$. (c) The extrapolated results for $\lambda (\beta)$ are compared with the theoretical prediction (dashed line); there is noticeable disagreement regardless of the fitting window specified by $F_1$ (in all cases $F_0 = 0.03$).
 }
        \label{fig:hyper_fit}
\end{figure}

As before, we perform least squares regression within a window defined by $F_0 \le F(t) \le F_1$. 
We then extrapolate the fitting parameter $\lambda_\textrm{fit}$ using a quadratic function in $1/N$ (the actual system size, not the fitting parameter $N_\textrm{fit}$). 
In Fig.~\ref{fig:hyper_fit}, we summarize the results of this approach.
In general, we find better agreement with theoretical predictions than with the previous exponential fits, especially at high temperatures. 
Upon closer inspection, however, it is evident that the fitted values for $\lambda$ do not extrapolate to the theoretical predictions, regardless of the choice for $F_0$ and $F_1$.
We conclude that this fitting procedure is not robust against finite-size and high-temperature corrections away from the low-temperature, semiclassical limit where it was derived. 

\subsection{Finite-size rescaling method}\label{sec:rescale}

\begin{figure}[t!]
  \includegraphics[width=0.8\linewidth]{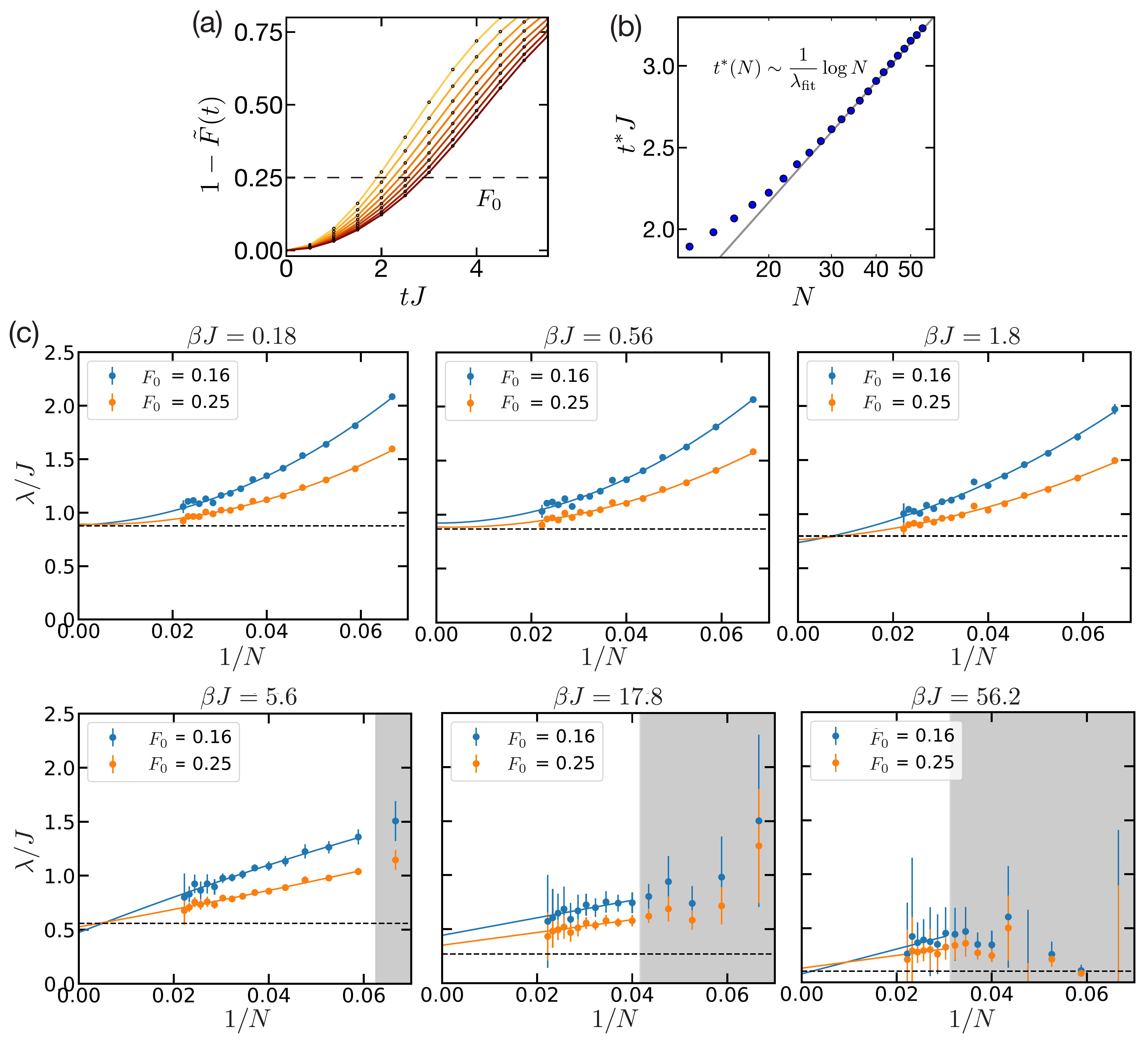}
  \caption{Finite-size rescaling procedure for extracting $\lambda$. (a) For a fixed temperature, we numerically simulate OTOCs across a range of system sizes. At each system size, the time $t^*(N)$ is determined for which the (normalized) OTOC, $\tilde F(t) \equiv F(t)/F(0)$, reaches a fixed value, i.e.~$\tilde F(t^*) = 1-F_0$. Data shown correspond to $\beta J = 1.8$ and $F_0 = 0.25$. (b) The Lyapunov exponent $\lambda$ is computed from the slope of $t^*(N)$ with respect to $\log N$ via an extrapolation procedure. (c) The extrapolation procedure is shown for each temperature and $F_0 = 0.16$ and $0.25$. We approximate $\lambda_\text{fit}(N)$ (data) by computing the finite difference $\Delta t^*(N)/\Delta(\log N)$ between successive system sizes. We then perform a $1/N$ extrapolation (solid lines) on the subset of system sizes (white area) whose first 20 eigenvalues are within $\Delta E = \beta J$ of the ground state. To avoid over-fitting, the extrapolation relies on a quadratic function for $\beta J \ge 5.6$ and a linear function for lower temperatures. 
 }
        \label{fig:fig_extrap}
\end{figure}

Having ruled out the possibility of fitting our data to a simple functional form, we now introduce a model-free method for extracting the Lyapunov exponent based on finite-size scaling. 
The only assumption we make is that the OTOCs approximately obey a rescaling symmetry of the form, 
\begin{equation} \label{scaling}
\begin{aligned}
&N \rightarrow r N \\
&t \rightarrow t + \frac 1 \lambda \log r
\end{aligned}
\end{equation}
This symmetry is expected to hold for any many-body chaotic model governed by ladder diagrams close to the semiclassical limit.

Based on this symmetry, we devise the following numerical procedure to extract $\lambda$. 
First, we compute the timescale at which the OTOCs reach a fixed value $\tilde F(t_N) = F_0$, for system size $N$; this requires interpolating our numerical data and solving for the intercept at $F_0$.
Second, we estimate $\lambda_\textrm{fit}(N)$ via a numerical derivative, i.e.~$1/\lambda_\textrm{fit}(N) = (t_N-t_{N-1})/(\log N - \log(N-1))$.
%
%
Finally, we extrapolate $\lambda_\textrm{fit}(N)$ to $N \rightarrow \infty$ using a polynomial extrapolation function, e.g.~$\lambda_\textrm{fit}(N) = \lambda_0 + \lambda_1/N + \lambda_2/N^2$.
The extrapolations for various temperatures are shown in Fig.~\ref{fig:fig_extrap}. 
%
%

A few comments are in order.
First, the extrapolation is performed on a subset of system sizes whose lowest 20 eigenstates lie within $\Delta E = 1/\beta$ of the ground state.
This criterion is meant to rule out systems that are dominated by the discreteness of the energy spectrum, for which no effective (replica-diagonal) action exists. 
Furthermore, to avoid overfitting, we use a quadratic extrapolation function for temperatures corresponding to $\beta \ge 5.6$ and a linear extrapolation for lower temperatures.
The reported error bars on our final results correspond to the standard error of the fitting parameter, $\lambda_0$.

Second, we note that the rescaling procedure depends on the value of $F_0$. 
In the large $N$ limit, the choice of this parameter is arbitrary, as the rescaling symmetry \eqref{scaling} is expected to hold for all values of $F_0$. 
At finite sizes, however, there are higher-order corrections that break the rescaling symmetry, particularly (i) at early times due to the microscopic cutoff, and (ii) at late times due to the crossover to power-law decay \cite{Bagrets_power-law_2017}. 
We thus expect an intermediate choice of $F_0$ to provide the best approximation. 
Our results in the main text are based on $F_0 = 0.25$. 
In Fig.~\ref{fig:exp_fit}, we show that a different choice, $F_0 = 0.16$, provides consistent results.
This demonstrates that our rescaling procedure is not overly sensitive to the precise value $F_0$.

\begin{figure}[t!]
  \includegraphics[width=\linewidth]{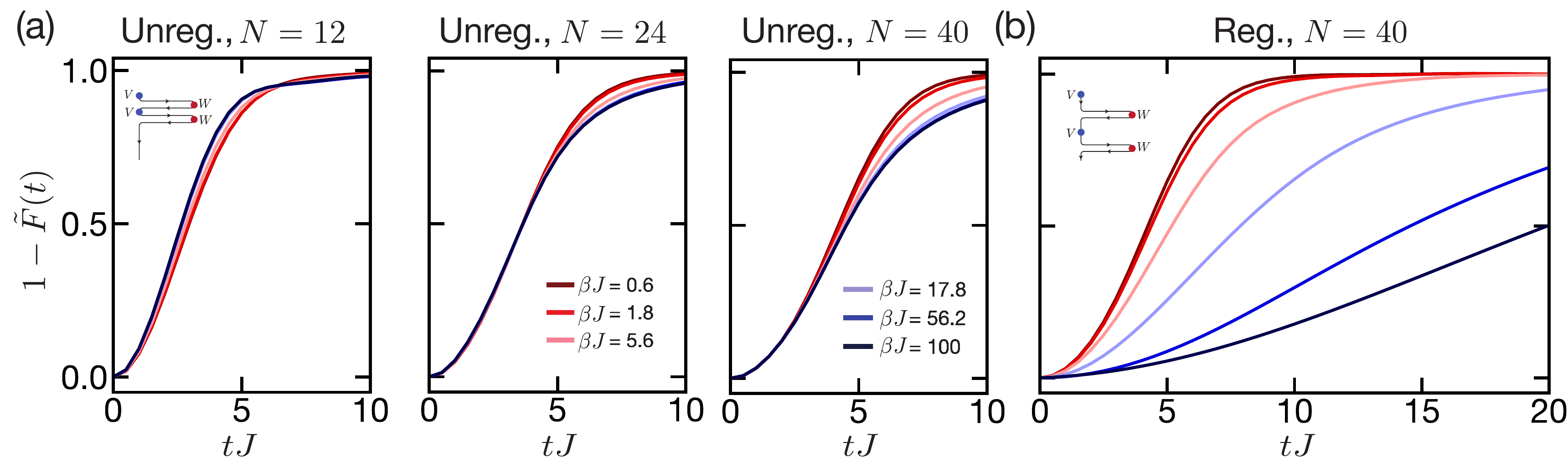}
  \caption{Numerical data for unregularized (a) and regularized (b) OTOCs at various temperatures. (a) The unregularized data correspond to the real part of the OTOC normalized by the initial value and is shown for three different system sizes: $N=12,24$ and $40$. For $N=12$ the lower temperature correlators grow faster than the high temperature correlators; this trend is reversed for $N\gtrsim 24$. (b) Compared to the unregularized data, the growth timescale for the regularized data increases significantly at low temperatures. This discrepancy is attributed to the difference in the scrambling time for the two types of correlators. In particular, the scrambling time for the unregularized OTOC is highly suppressed at low temperatures, implying that the observed growth in the numerics arises from dissipative dynamics rather than to chaos. (Inset) Schematic of the two configurations (unregularized and regularized), represented as a path in real (horizontal) and imaginary (vertical) time.
 }
        \label{fig:Fs_reg_vs_unreg}
\end{figure}

\begin{figure}[t!]
  \includegraphics[width=0.7\linewidth]{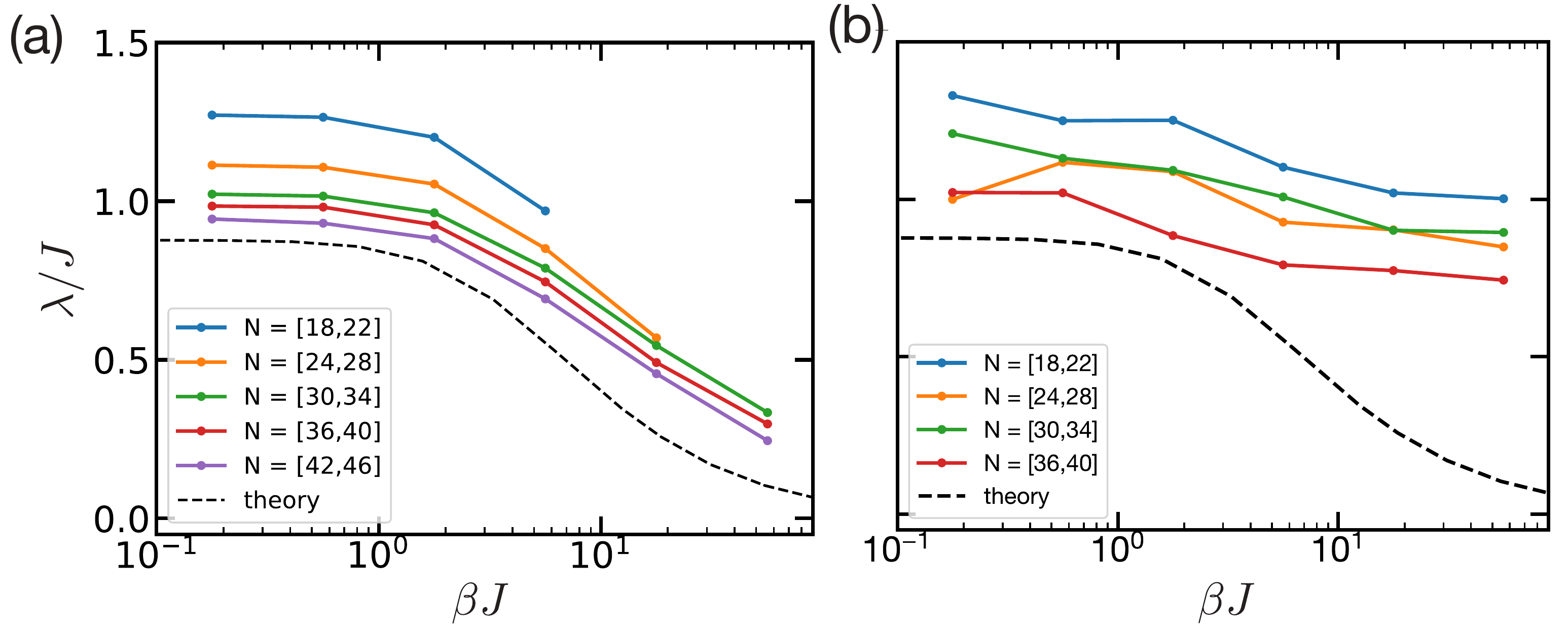}
  \caption{The finite-size rescaling procedure applied to  regularized (a) and unregularized (b) OTOCs. The data points corresponds to $\lambda_\textrm{fit}(N)$ for various size intervals and the dashed line is the theoretical prediction. The unregularized results exhibit a weaker temperature dependence, and there is a larger discrepancy from theoretical predictions at low temperatures.
 }
        \label{fig:lambda_reg_vs_unreg}
\end{figure}
\subsection{Unregularized OTOCs}\label{sec:unreg}
In the main text and in the results presented so far, we have considered a thermally regularized OTOC, given by
\begin{equation} \label{F_reg}
F^{(\textrm{r})}(t) \equiv \overline {\left < \chi_i(t) \rho^{\frac 1 4} \chi_j(0) \rho^{\frac 1 4} \chi_i (t) \rho^{\frac 1 4} \chi_j (0) \rho^{\frac 1 4} \right > }.
\end{equation}
This form of the correlator has the advantage that it is Hermitian and can be analyzed at low temperatures with the effective low-energy action (without needing additional regularization). 
However, for many experiments, it is more convenient to measure the unregularized OTOC, given by
\begin{equation} \label{F_reg}
F^{(\textrm{u})}(t) \equiv \overline {\left < \chi_i(t) \chi_j(0)\chi_i (t) \chi_j (0) \rho \right > }.
\end{equation}
%
Here, we provide numerical evidence that the unregularized correlator is subject to stronger finite-size corrections at low temperatures, a claim that is supported theoretically in later sections \footnote{Similar numerical data showing larger finite-size corrections for the unregularized correlators was also presented in \cite{lantagne2020diagnosing}.}.

In Fig.~\ref{fig:Fs_reg_vs_unreg}(a), we present numerical data for the unregularized correlator at three system sizes: $N=12,24$ and $40$.
As mentioned in the main text, the correlator at the smallest size exhibits the \emph{opposite} temperature dependence as expected, i.e.~OTOCs at low temperatures appear to grow \emph{faster} than at high temperatures.
This trend is consistent with previous studies based on exact diagonalization \cite{Fu_numerical_2016}.
At $N = 24$ we observe an inversion in this trend, and thereafter the correlator exhibits the expected qualitative temperature dependence.

However, even at the largest system size, the temperature dependence of the unregularized correlator is significantly weaker than the regularized correlator [Fig.~\ref{fig:Fs_reg_vs_unreg}(b)].
This difference can be quantified by repeating our finite-size scaling analysis with the unregularized correlator (Fig.~\ref{fig:lambda_reg_vs_unreg}).
For temperatures above $\beta J \approx 10$, regularization has little effect on $\lambda_\textrm{fit}(N)$, i.e.~the estimate for $\lambda$ at size $N$. 
However, for lower temperatures, the values for $\lambda_\textrm{fit}(N)$ using the unregularized correlator show a weaker temperature dependence than in the regularized case and are substantially farther from the theoretical predictions \footnote{We emphasize that $\lambda_\textrm{fit}(N)$ should not be interpreted as a size-dependent Lyapunov exponent, but rather as a fitting parameter that can be extrapolated to $N\rightarrow \infty$ to determine the Lyapunov exponent.}.
This implies that the unregularized correlator is subject to more pronounced finite-size effects, and larger system sizes are required to accurately determine the Lyapunov exponent.

We attribute these results to the fact that the scrambling time is shorter for the unregularized correlator and, hence, there is less separation between the early-time dissipative dynamics and the chaotic growth (see Sec.~\ref{sec:unreg} and \ref{sec:Sch_OTOC}).
Such considerations are important for numerical and experimental studies of OTOCs; however, we emphasize that they do not imply that the two forms of the correlator are characterized by distinct Lyapunov exponents.
Indeed, we justify mathematically in Sec.~\ref{sec:reg_vs_unreg} that the Lyapunov exponent (for the SYK model) is independent of regularization.
%

\section{Large $N$ solutions}

\begin{figure}[b!]
  \includegraphics[width=0.4\linewidth]{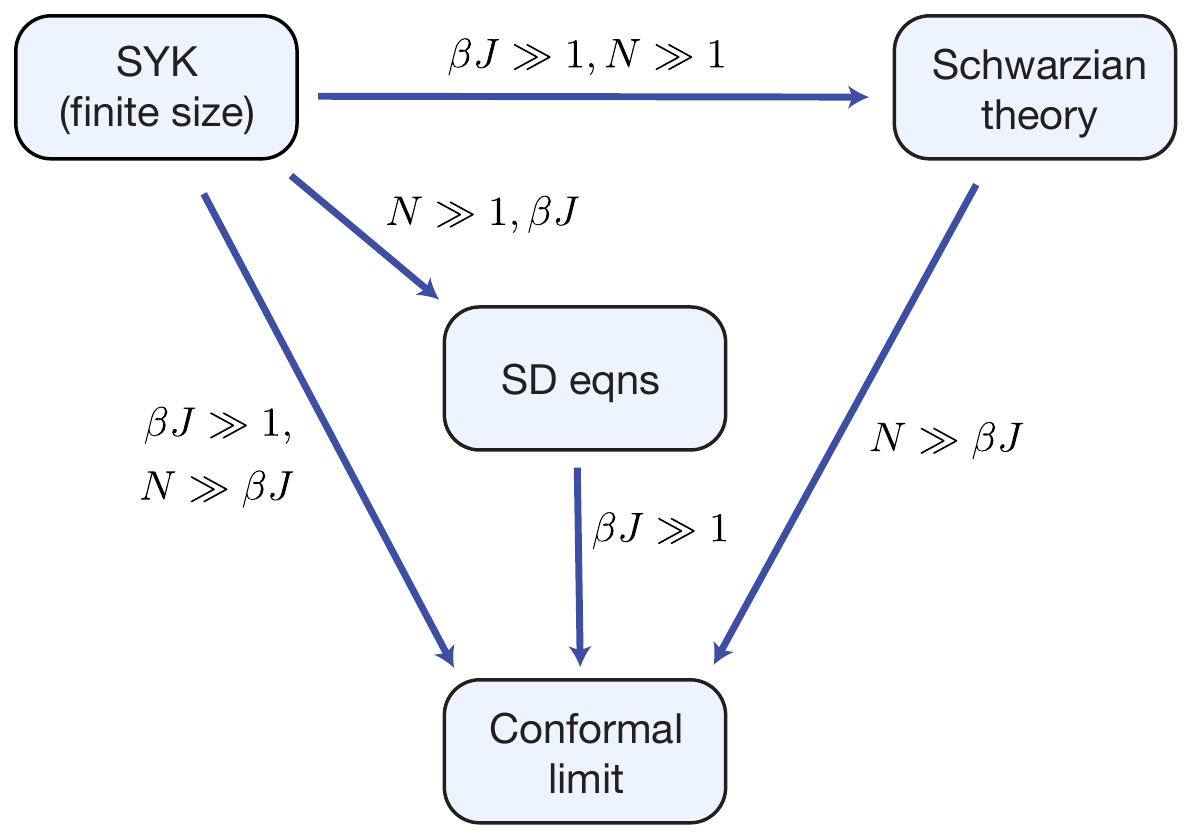}
  \caption{Schematic relationship between various theories describing the dynamics of the SYK model. The Schwinger-Dyson (SD) equations represent the large $N$ solution at all temperatures \eqref{SD}, while the Schwarzian action \eqref{Schwarzian} captures finite $N$ behavior at low temperatures. In the limit $N \gg \beta \gg 1$, both theories approach the conformal limit \eqref{conformal}.
 }
        \label{fig:flow_chart}
\end{figure}

In this section, we describe the large $N$, semiclassical solutions for the dynamics of the SYK model.
These results were derived previously via either a diagrammatic approach or from the saddle-point of a disorder averaged effective action \cite{Sachdev_gapless_1993,Kitaev_simple_2015,Maldacena_remarks_2016,Bagrets_liouville_2016}.
Finite-size corrections beyond the semiclassical solution can be computed at low temperatures via the Schwarzian action and are discussed in Sec.~\ref{sec:Schwarzian}.
In Fig.~\ref{fig:flow_chart}, we illustrate schematically the relationship between these various theories.

%

\subsection{Schwinger-Dyson equations}

The large $N$ solution for the average Green's functions is given in imaginary time by the self-consistent Schwinger-Dyson equations:
\begin{equation} \label{SD}
\frac 1 {G(\omega)} = i \omega + \Sigma(\omega) \; ,\quad \Sigma(\tau) = J^2 \left [ G(\tau)^3\right ],
\end{equation}
where $\omega$ are the Fourier components with respect to $\tau$.
The real-time version of these equations is obtained by setting $\tau = i t$.
At low temperatures ($\beta J \gg 1$) and long timescales ($\tau J ,t J \gg 1$), the derivative term $i\omega$ can be neglected, leading to an emergent conformal symmetry.
The solution in this limit is \cite{parcollet1999non,Maldacena_remarks_2016}
\begin{equation} \label{conformal}
G_c(\tau) = b \left [\frac \pi {\beta \sin{\frac{\pi \tau}{\beta}}} \right ]^{1/2}, \quad G_c(it) = b \left [\frac \pi {i \beta \sinh{\frac{\pi t}{\beta}}} \right ]^{1/2}, 
\end{equation}	
where $b = 1/(\sqrt{2}\pi^{1/4}) \approx 0.531$.
More generally, the Green's functions can be computed at all temperatures by solving \eqref{SD} through an iterative numerical approach \cite{Maldacena_remarks_2016}.
This procedure converges quickly in imaginary time and yields the results shown in Fig.~3(a) of the main text.
In real time, the numerical analysis is more subtle, and we found that the most stable approach is the implementation proposed in \cite{Eberlein_quantum_2017}.
We present these results in Fig.~3(b) of the main text and in Fig.~\ref{fig:GR} below.  
We also rely on the real-time correlators to compute the Lyapunov exponent and magnitude of OTOCs, as described in the following section.
For both real and imaginary time, we benchmarked the numerical solutions by comparing to \eqref{conformal} at low temperatures.

\subsection{Kernel equation}

The leading order behavior for OTOCs is computed via a set of diagrams known as ladder diagrams. 
In particular, one defines
\begin{equation} \label{F_growth}
F(t_1,t_2) \equiv \left < \chi_i(t_1) \chi_j(0) \chi_i(t_2) \chi_j(0) \right > = \mathcal{F}_0 + \frac 1 N \mathcal{F}(t_1,t_2) 
\end{equation}
and makes a growth ansatz of the form
\begin{equation}
\mathcal{F}(t_1,t_2) = e^{\lambda (t_1+t_2)/2} \gamma(t_{12}) 
\end{equation}
where $t_{12} = t_2-t_1$.
The exponent is determined by solving the eigenvalue equation
\begin{equation} \label{kernel_eqn}
\mathcal{F}(t_1,t_2) = \int dt_3 dt_4 K_R(t_1,t_2,t_3,t_4) \mathcal{F}_1(t_3,t_4)
\end{equation}
with eigenvalue one. 
Here $K_R$ is the retarded kernel given by
\begin{equation} \label{KR}
K_R(t_1,t_2,t_3,t_4) = J^2(q-1)G_R(t_{13})G_R(t_{24})G_W(t_{34})^{q-2}
\end{equation}
where $G_R(t)$ is the retarded Green's function, $G_W(t)$ is known as the Wightman function, and $q=4$ is the number of fermions involved in each interaction in the SYK Hamiltonian (from here on, we generalize to $q$-body interactions).
Crucially, the Wightman function depends on the regularization. 
In the case of the regularized OTOC, the Wightman function is given by $G_W(t) = G(t+i\beta/2)$ and can be computed numerically from the Schwinger-Dyson equations.

To determine the Lyapunov exponent (for the regularized OTOC), we perform the following numerical procedure.
First, we solve for the eigenvalues of $K_R$ for a given value of $\lambda$.
This relies on the numerical results for $G_R(t)$ and $G_W(t)$ and the discretization of time into $M$ steps.
Second, we perform a binomial search to find $\lambda$ corresponding to a maximum eigenvalue of one.
Finally, we repeat the procedure with different values for $M$ and extrapolate to estimate $\lambda$ in the continuous limit. 
The numerical results for $\lambda$ are shown in Fig.~1(b) of the main text; we verify that the results agree with the low-temperature limit, $\lambda \approx 2\pi/\beta$.

\subsubsection{Regularized vs. unregularized exponent}\label{sec:reg_vs_unreg}

In principle, one can obtain the Lyapunov exponent for the unregularized OTOC in an analagous way: by calculating the Wightman function with no imaginary-time separation, i.e.~$G_W(t) = G(t)$, and repeating the numerical procedure outlined above.
However, the numerical analysis is more challenging, as the kernel is no longer Hermitian; in particular, we found that the eigenvalues are highly sensitive to numerical errors that arise from discretization and the imprecision of the Green's functions. 

Nevertheless, one can show theoretically that the Lyapunov exponent is independent of regularization (for the SYK model or similar ladder-diagram theories).
To do so, we begin by defining the kernel ansatz in the regularized case as
\begin{equation} \label{kernel_reg}
e^{\lambda (t_1+t_2)/2} \gamma^{(r)}(t_{12}) = J^2(q-1)\int dt_3dt_4 G_R(t_{13})G_R(t_{24})G_W^{(r)}(t_{34})^{q-2}e^{\lambda (t_3+t_4)/2} \gamma^{(r)}(t_{34})
\end{equation}
and in the unregularized case as
\begin{equation} \label{kernel_unreg}
e^{\lambda (t_1+t_2)/2} \gamma^{(u)}(t_{12}) = J^2(q-1)\int dt_3dt_4 G_R(t_{13})G_R(t_{24})G_W^{(u)}(t_{34})^{q-2}e^{\lambda (t_3+t_4)/2} \gamma^{(u)}(t_{34})
\end{equation}
where $G^{(r)}_W(t) = G(t+i \beta/2)$ and $G^{(u)}_W(t) = G(t)$ and $\gamma^{u,r}(t_{12})$ is the normalizable eigenvector with respect to inner product:
\begin{equation}
	(\gamma,\gamma)_{r,u}=(q-1) J^2\int_{-\infty}^{\infty}dt \gamma(t)G_W^{(r,u)}(t)^{q-1}\gamma(t).
\end{equation} 

The proof is then as follows \footnote{We note that during the preparation of this manuscript a similar proof was presented in \cite{RomeroBermudez_regularization_2019}.}.
By definition, \eqref{kernel_unreg} can be written in terms of the $G^{(r)}_W(t)$ as
\begin{equation} 
e^{\lambda (t_1+t_2)/2} \gamma^{(u)}(t_{12}) = J^2(q-1)\int dt_3dt_4 G_R(t_{13})G_R(t_{24})G_W^{(r)}(t_{34}-i\beta/2)^{q-2}e^{\lambda (t_3+t_4)/2} \gamma^{(u)}(t_{34}).
\end{equation}
We next reparameterize the time variables as $t_1^\prime = t_1 + i\beta/4$, $t_2^\prime = t_2 - i\beta/4$, $t_3^\prime = t_3 + i\beta/4$, and $t_4^\prime = t_4 - i\beta/4$.
Using the fact that the integrand goes to zero at $t_{3,4}\rightarrow \pm \infty$, we can deform the integration contour and get:
\begin{equation} 
e^{\lambda (t_1^\prime+t_2^\prime)/2} \gamma^{(u)}(t_{12}^\prime+i\beta/2) = J^2(q-1)\int dt_3^\prime dt_4^\prime G_R(t_{13}^\prime)G_R(t_{24}^\prime)G_W^{(r)}(t_{34}^\prime)^{q-2}e^{\lambda (t_3^\prime+t_4^\prime)/2} \gamma^{(u)}(t_{34}^\prime+i\beta/2).
\end{equation}
In addition, using the fact that $\gamma^r(t)$ is normalizable, one can show that $\gamma^{r}(t+i\beta/2)$ is also normalizable with respect to the inner product for the unregularized case. We thus recover the regularized equation \eqref{kernel_reg} and identify the relation $\gamma^{(r)}(t) = \gamma^{(u)}(t+i\beta/2)$.

In summary, changing the regularization has no effect on the growth exponent but only on the the eigenfunction $f(t)$ (which controls the magnitude of OTOCs, as shown below).
More generally, this argument applies to any degree of regularization, i.e.~$G_W(t) = G(t+i\eta)$, and to any other many-body chaotic model whose OTOCs are described by ladder diagrams.


\subsection{Magnitude of OTOCs}\label{sec:mag}
The magnitude of the growing term of the OTOC is important because it determines the scrambling time and, thus, the separation between early-time dissipative dynamics and chaotic growth.
In particular, we define the magnitude of exponential growth $C_1$ as
\begin{equation} \label{eq:mag}
\tilde F(t) \equiv \frac{F(t)}{F(0)} \approx 1 - \frac {C_1} N e^{\lambda t} + \mathcal O ({1/N^2})
\end{equation}
Having a well-separated scrambling time, $t^* \approx 1/\lambda \log (N/C_1)$, then corresponds to $N \gg C_1$.

Recently, Gu and Kitaev derived an identity that relates the magnitude of the leading order growth term in OTOCs with other quantities computable from the Schwinger-Dyson equations and the kernel equation. 
The identity is given by
\begin{equation} \label{magnitude}
\frac 1 {C_1} = 2 \cos\left( \frac {\lambda \beta}{4} \right) t_B \left(\gamma,\gamma\right)\end{equation}
where $t_B = k^\prime(\lambda)$ is the ``branching time'', $k$ is the eigenvalue of the retarded kernel, \eqref{KR}, and $(\gamma,\gamma)$ is given by
\begin{equation}
(\gamma,\gamma) \equiv (q-1)J^2 \int_{-\infty}^\infty dt \gamma(t)\left(G_W(t)\right)^{q-2}\gamma(t)
\end{equation}
We note that $\gamma(t)$ and $G_W(t)$ are defined in the previous section, and both depend on the choice of regularization.

\begin{figure}[t!]
  \includegraphics[width=0.4\linewidth]{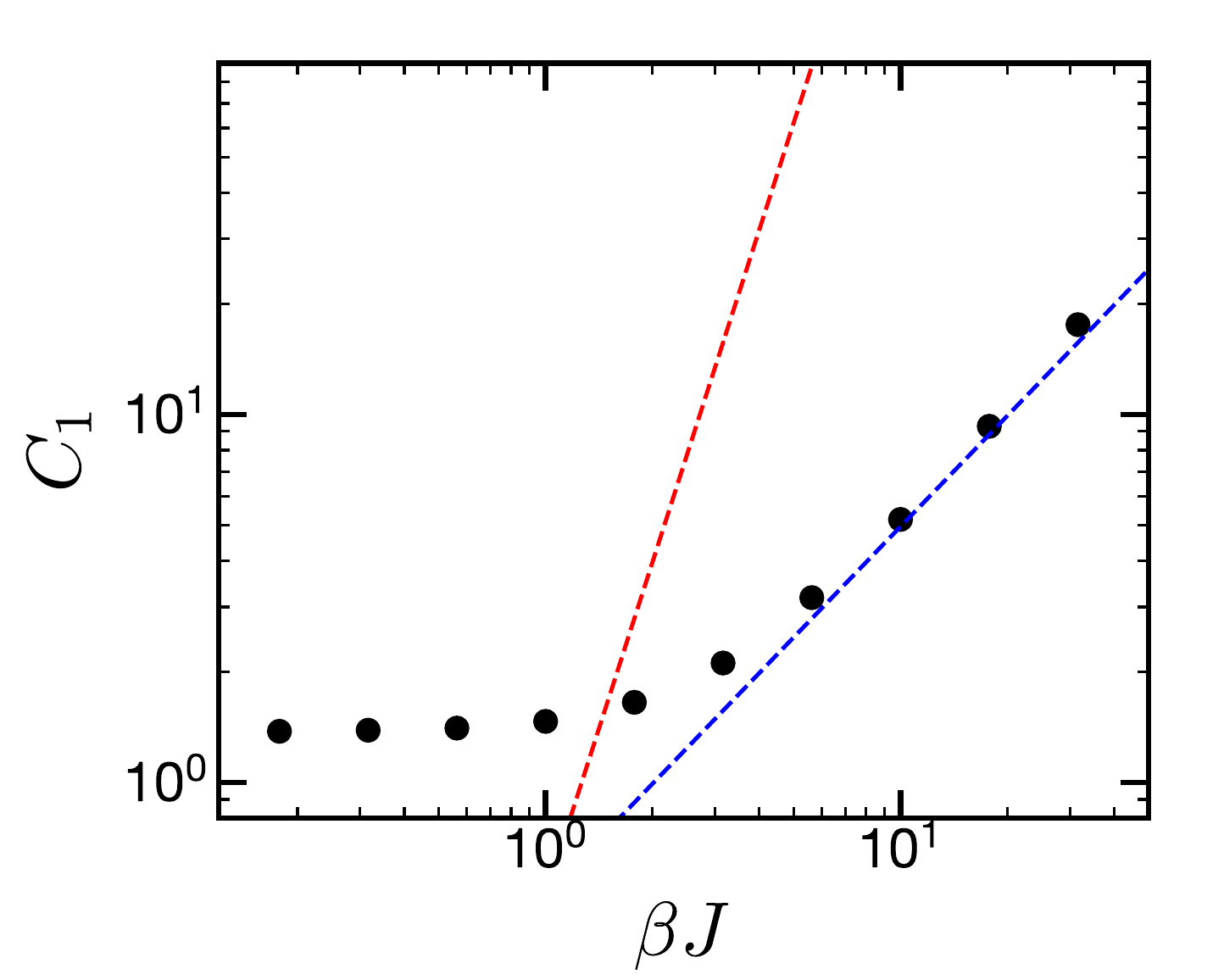}
  \caption{The normalized magnitude, $C_1$, of the leading-order term in the OTOC, $C_1 / N e^{\lambda t}$, as a function of temperature. The magnitude is calculated numerically (black) for the regularized OTOC using \eqref{magnitude}; at high temperatures $C_1 \approx 1.4$ and at low temperatures $C_1 \approx 0.5 \beta J$, in agreement with the semiclassical solution, \eqref{F_Sch} (blue). For the unregularized OTOC, the semiclassical solution predicts the low-temperature scaling $C_1 \sim (\beta J)^3$ (red).
 }
        \label{fig:mag}
\end{figure}

For the regularized OTOC, we solve for $C_1$ as a function of temperature using \eqref{magnitude} and the numerical solution of the kernel equation.
These results are shown in Fig.~\ref{fig:mag}.
At high temperatures $C_1$ approaches $1.4$, while at low temperatures $C_1 \approx 0.5 \beta J$.  
This latter result is consistent with previous work \cite{Maldacena_remarks_2016,Gu_relation_2019} and provides validation of our numerical methods. 

For the unregularized OTOC, the magnitude can in principle be calculated following the same approach; however, we confront the same numerical difficulties regarding the kernel equation as discussed in Sec.~\ref{sec:reg_vs_unreg} and thus leave this computation for a future work. 
Of course, at high temperatures ($\beta J \ll 1$), the magnitude must be the same regardless of regularization.
At low temperatures ($\beta J \gg 1$), we can use the Schwarzian action to determine the scaling $C_1 \sim (\beta J)^3$, as discussed in Sec.~\ref{sec:Sch_OTOC}.

In combination, these results imply that having a separation of timescales is achieved in the regularized case when $N \gg \beta J$ and in the unregularized case when $N \gg (\beta J)^3$.
As the latter condition is more stringent, it provides an explanation for the more severe finite-size effects observed with the unregularized correlator (see Sec.~\ref{sec:unreg}).

\section{Schwarzian action}\label{sec:Schwarzian}

In the previous section, we focused on the semiclassical solution of the SYK dynamics. 
We now discuss corrections about this limit obtained via the Schwarzian action. 
This 0+1 dimensional action is valid for describing the SYK model at low energies (i.e.~$\beta J, \tau J \gg 1$), and is given by
\begin{equation} \label{Schwarzian}
S_\textrm{Sch} = -C_\textrm{Sch} \int_0^\beta d\tau \left\{f,\tau\right\}, \quad \left\{f,\tau\right\}, \equiv \frac{f^{'''}}{f^{'}} - \frac 3 2 \left(\frac{f^{''}}{f^{'}}\right)^2
\end{equation}
where $f(\tau)$ is a function describing reparameterizations of time, i.e.~$\tau \rightarrow f(\tau)$.
Prior work has established the relation between the coupling coefficient $C_\textrm{Sch}$ and parameters of the SYK model: $C_\textrm{Sch} \approx 0.01 N/J$ \cite{Maldacena_remarks_2016}\footnote{We note that there is some disagreement about this relation; in particular, some studies suggest $C_\textrm{Sch} = 1/(64 \pi^{1/2} J)\log J/\Delta$ \cite{Bagrets_liouville_2016}. Our numerics for two-point functions support the original relation derived in \cite{Maldacena_remarks_2016}}. 
This prefactor controls the size of fluctuations about the $C_\textrm{Sch} \rightarrow \infty$ saddle-point solution.

Recently, analytical methods have been developed to solve the full dynamics of the Schwarzian action, enabling the calculation of correlation functions at all orders in $1/C_\textrm{Sch}$  \cite{Bagrets_liouville_2016,Lam_shockwave_2018,Yang_quantum_2019}.
Furthermore, these developments have established a direct correspondence between the Schwarzian action and near AdS in 1+1 dimensions \cite{Kitaev_soft_2018}. 
In particular, the saddle-point solution is dual to classical gravity, while higher order $1/C_\textrm{Sch}$ corrections correspond to gravitational fluctuations. 

\subsection{Two-point correlators}

We consider the two-point function, $G(z) = \left < \chi_i(z) \chi_i(0) \right >_\beta$, where $z=it+\tau$ is complexified time and the thermal average is computed at inverse temperature $\beta$. 
Using the Schwarzian action, the exact result for $G(z)$ can be computed and is given by \cite{Bagrets_liouville_2016,Lam_shockwave_2018,Yang_quantum_2019}
\begin{equation} \label{G_Sch}
G_\textrm{Sch}(z) = \frac 1 {\mathcal{N} (2C_\text{Sch})^{2\Delta} \Gamma(2\Delta)} \int ds_1 ds_2 \rho(s_1)\rho(s_2)e^{-\frac{s_1^2}{2C}z-\frac{s_2^2}{2C}(\beta-z)}\left|\Gamma(\Delta-i(s_1+s_2))\Gamma(\Delta+i(s_1-s_2))\right|^2
\end{equation}
where $\rho(s)=\frac{s}{2\pi^2}\sinh(2\pi s)$ is the density of states, $\Gamma(\cdot)$ is the Gamma function, and the normalization factor is equal to
\begin{equation}
\mathcal{N} = \frac {C_\text{Sch}^\frac 3 2} {\sqrt{2\pi} \beta^{\frac 3 2}} e^{\frac {2 C_\text{Sch} \pi^2} \beta}.
\end{equation}

\begin{figure}[b!] 
  \includegraphics[width=\linewidth]{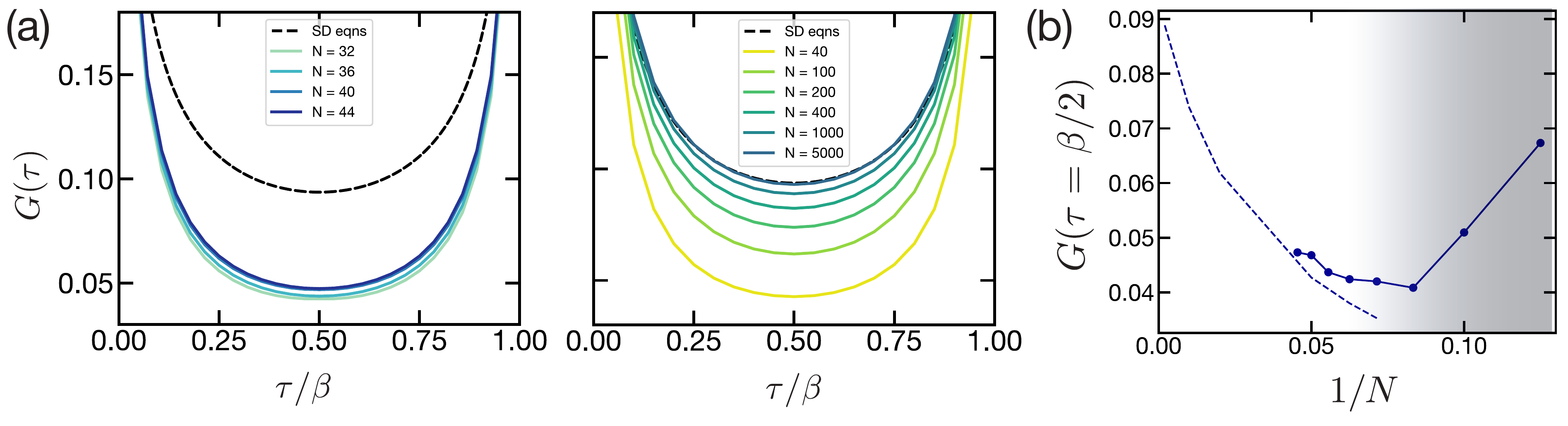}
  \caption{(a) Results for the imaginary-time Green's function, $G(\tau)$, at $\beta J =100$ from our numerics (left) and the solution of the Schwarzian action (right). (b) Finite-size scaling of $G(\tau = \beta / 2)$. Our numerics (data) begin to agree with the Schwarzian solution (dashed line) at $N\approx 30$. For smaller sizes (gray area), we observe finite-size effects that cannot be accounted for by the Schwarzian action, which we attribute to the discreteness of the energy spectrum. 
 }
 \label{fig:G_imag}
\end{figure}

The behavior of $G(z)$ can be understood qualitatively in several regimes. 
At short times $t \ll C_\textrm{Sch}$, the integrals are well-approximated by the classical saddle point, leading to the aforementioned conformal solution \eqref{conformal}.
In contrast, at late times $t \gg C_\textrm{Sch}$, the behavior is dominated by the low-energy edge of the spectrum, which respects the Wigner semicircle law: $\rho(E) \sim \sqrt{E}$ or $\rho(s) \sim s^2$. 
This gives rise to a power-law decay with an exponent independent of the operator dimension.
In particular, we can identify two cases, depending on the temperature relative to $C_{\textrm{Sch}}$:
\begin{enumerate}[label=(\alph*)]
\item High temperature, $\beta \ll C_\textrm{Sch}$:
\begin{equation}{}
G_\textrm{Sch}(t) \sim \begin{cases}
    G_c(t), & t \ll C_\textrm{Sch} \\
    t^{-3}, & t \gg C_\textrm{Sch} 
  \end{cases}
\end{equation}
\item Low temperature, $\beta \gg C_\textrm{Sch}$:
\begin{equation} \label{eq:G_low_T}
G_\textrm{Sch}(t) \sim \begin{cases}
    G_c(t), & t \ll C_\textrm{Sch} \\
    \frac{\beta^{3/2}}{(i t(\beta - i t))^{3/2}} , & t \gg C_\textrm{Sch} 
  \end{cases}
\end{equation}
\end{enumerate}

\begin{figure}[t!]
  \includegraphics[width=\linewidth]{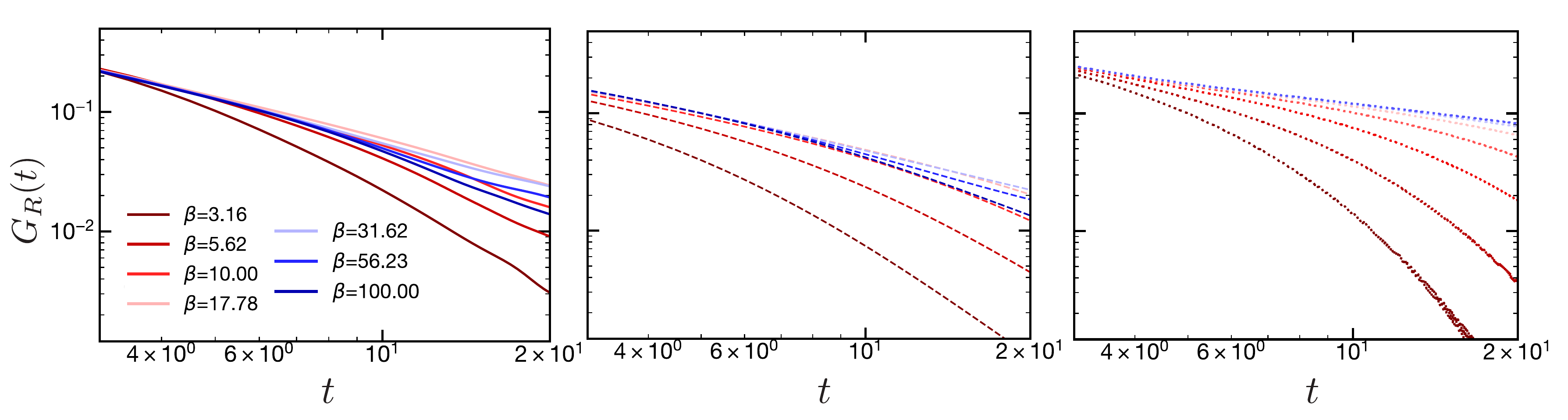}
  \caption{The retarded Green's function for the SYK model with 40 Majoranas (left), the Schwarzian action with $C_\text{Sch}=0.4$ (middle), and the large $N$ Schwinger-Dyson equations (right). The late-time behavior of of the SYK results and the Schwarzian action is governed by the random-matrix-like form of the energy spectrum, which manifests as a non-monotonicity with respect to temperature. 
 }
        \label{fig:GR}
\end{figure}

While observing a clear separation between these regimes is challenging at finite-sizes, we expect our numerical results to be described quantitatively by the full functional form of $G_\textrm{Sch}(z)$. 
To this end, we compute the integrals in \eqref{G_Sch} numerically in Mathematica.  
The integrals converge quickly for the imaginary-time correlator, $G_\textrm{Sch}(\tau)$; for the real-time correlator, obtaining convergence requires us to introduce a small imaginary-time separation, i.e.~$G_\textrm{Sch}(it) \rightarrow G_\textrm{Sch}(it+\epsilon)$.
The numerical results in imaginary and real time are shown in Fig.~\ref{fig:G_imag} and \ref{fig:GR}, respectively, for the temperatures and timescales relevant to our study.
We note that the retarded Green's function corresponds to the real part of $G_{\textrm{Sch}}$.
A key feature of this correlator is a non-monotonic decay with respect to temperature; this results from the non-trivial dependence of the \emph{phase} on $t$ and $\beta$ in \eqref{eq:G_low_T} and is in stark contrast with the prediction of the conformal solution or of the semiclassical solution of the SYK model (i.e.~the Schwinger-Dyson equations). 


\subsection{Out-of-time-order correlators}\label{sec:Sch_OTOC}

While previous studies have derived an exact integral expression for the OTOC analogous to \eqref{G_Sch}, the integrals are significantly more complex and solving them numerically is beyond the scope of this study \cite{Lam_shockwave_2018}. 
Instead, we rely on the semiclassical limit of the OTOC, which is valid for $C_\textrm{Sch} \rightarrow \infty, t \lesssim 1/\lambda \log C_\textrm{Sch}$. 
This expression was derived using the correspondence to quantum gravity (i.e.~by summing over tree-level graviton diagrams) and is given by \cite{Maldacena_conformal_1606}
\begin{equation} \label{F_Sch}
\frac {F(z_1,z_2,z_3,z_4)} {F(\tau_1,\tau_2,\tau_3,\tau_4)} = \frac {U(2\Delta,1,\frac 1 z)}{z^{2\Delta}}, \quad z = \frac {i\beta} {16 \pi C_\textrm{Sch}} \frac {e^{\frac {2\pi}{\beta}(z_3+z_4-z_1-z_2)/2}}{\sinh \frac {\pi z_{12}} {\beta} \sinh \frac {\pi z_{34}} {\beta}}
\end{equation}
where $z_i = t_i + i\tau_i$ are the complexified times at which the four operators are inserted, i.e.~$\left < V(z_1) W(z_3)V(z_2)W(z_4)\right >$, and $U(\cdot)$ is the confluent hypergeometric function.

From \eqref{F_Sch}, we can determine the behavior of different regularizations.
The regularized OTOC corresponds to 
\begin{equation}
z_1 = i \frac \beta 2, \; z_2 = i \frac \beta 2, \; z_3 = t-i \frac \beta 4, \; z_4 = t+i \frac \beta 4
\end{equation}
which leads to
\begin{equation}
\frac {F(t)}{F(0)} = \frac {U(2\Delta,1,\frac 1 z)}{z^{2\Delta}} , \quad z = \frac {\beta}{16 \pi C_{\textrm{Sch}}} e^{\frac {2\pi} \beta t}
\end{equation}
Crucially, this expression can be expanded in powers of $e^{\lambda t}/C_\textrm{Sch}$. For example, setting $\Delta = 1/4$, leads to
\begin{equation}
\frac {F(t)}{F(0)} = 1 - \frac {\beta}{64 \pi }\left(\frac{ e^{\lambda t}}{C_\textrm{Sch}} \right ) + \frac {9 \beta^2}{8192 \pi }\left(\frac{ e^{\lambda t}}{C_\textrm{Sch}} \right )^2 + \cdots
\end{equation}
As $C_\textrm{Sch} \sim N$, this manifestly satisfies the rescaling ansatz discussed in Sec.~\ref{sec:rescale} and the main text.
Moreover, the magnitude of the leading order term 
\begin{equation}
\frac {F(t)}{F(0)} \approx 1 - \frac {0.5 \beta J}{N} e^{\lambda t}
\end{equation}
agrees with the numerical results determined numerically in Sec.~\ref{sec:mag}.

To obtain different regularizations, we can reduce the imaginary-time separation between the operators. 
For example, setting
\begin{equation}
z_1 = i 2\eta, \; z_2 = i 2\eta, \; z_3 = t-i \eta, \; z_4 = t+i \eta
\end{equation}
corresponds to a symmetric separation with energy scale $\eta$.
It is evident that any finite value for $\eta$ leads to a well-defined OTOC with the same Lyapunov exponent as the fully regularized case.
To represent the fully unregularized correlator, the naive expectation is to take the limit $\eta \rightarrow 0$, which causes the denominator in \eqref{F_Sch} to vanish. 
For the SYK model, this UV divergence is clearly unphysical, and one should instead impose a microscopic cutoff of order $J$. 
The net effect is to enhance to growth term by a factor of $(\beta J)^2$ and thus decrease the scrambling time by a factor of $\log (\beta J)^2$. 
More precisely, the leading order term for the unregularized correlator is given by
\begin{equation}
\frac {F(t)}{F(0)} \approx 1 - c_1 \frac {(\beta J)^3}{N} e^{\lambda t}
\end{equation}
where $c_1$ is an order one prefactor.
The exact numerical value of $c_1$ cannot be determined by these methods, as the microscopic cutoff corresponds to ``smearing'' the operators over an energy scale $J$ rather than setting an exact value for $\eta$.  

To summarize, regularization changes the magnitude of the exponential growth but has no effect on the Lyapunov exponent. 
The intuition behind this conclusion can be understood from the dual gravitational theory, where the Lyapunov exponent corresponds to the coupling strength of the graviton interaction and the regularization corresponds to the initial energy of an incoming shock wave.
While the energy of the initial state has no effect on the coupling strength, it determines the timescale at which nonlinear graviton effects become relevant, leading to the saturation of the correlator.
In particular, the unregularized correlator corresponds to a higher energy initial state, which reaches saturation at an earlier time.
While this intuition applies directly to the SYK model at low temperatures, we expect the same qualitative effects to hold at all temperatures due to the form of the ladder diagrams; in the general case, the graviton interaction would be replaced by a `scramblon' interaction whose strength is governed by $\lambda(\beta)$.

\bibliography{ref_manual.bib,ref_auto.bib}